\documentclass[a4paper,11pt]{article}
\pdfoutput=1 

\usepackage{jcappub} 

\usepackage[T1]{fontenc} 
\usepackage{datetime}
\usepackage{amsmath}
\usepackage{amsfonts}
\usepackage{amssymb}
\usepackage{latexsym}
\usepackage{graphicx}
\usepackage{color}
\usepackage{mathrsfs}
\usepackage{slashed}
\usepackage{hyperref}
\usepackage{enumerate}
\usepackage[table]{xcolor}
\usepackage{colortbl}
\usepackage{multirow}
\usepackage{url}
\usepackage{enumitem}
\usepackage{comment}
\usepackage{yfonts}
\usepackage{tikz}
\usepackage{bm}
\usepackage{geometry}
\usepackage{lscape}
\usepackage{pgfgantt}
\usepackage{lipsum}
\usepackage{eurosym}
\usepackage{amsthm}
\definecolor{VioletRed4}{rgb}{0.55, 0.13, .32}

\hypersetup{colorlinks, citecolor=bluscuro, linkcolor=black, urlcolor=bluscuro}
\definecolor{rossos}{cmyk}{0,1,1,0.55}
\definecolor{bluscuro}{rgb}{0.15, 0.2, .85}
\definecolor{bluchiaro}{cmyk}{1,.3,0.,0.1}
\usepackage{cancel}
\def\hhref#1{\href{http://arxiv.org/abs/#1}{#1}} 
\definecolor{oucrimsonred}{rgb}{0.6, 0.0, 0.0}
\definecolor{persianblue}{rgb}{0.11, 0.22, 0.73}
\definecolor{forestgreen}{rgb}{0.13,0.35,0.13}
 \hypersetup{colorlinks, citecolor=oucrimsonred, linkcolor=persianblue, urlcolor=oucrimsonred}
 \hypersetup{colorlinks, citecolor=oucrimsonred, linkcolor=persianblue, urlcolor=oucrimsonred}

\definecolor{Gray}{gray}{0.95}
\definecolor{Grayyy}{gray}{0.95}

\newcommand{\fd}[2]{\parbox{#1}{\includegraphics[width=#1]{figs/#2}}}

\numberwithin{equation}{section}

\hypersetup{colorlinks, citecolor=persianblue, linkcolor=VioletRed4, urlcolor=forestgreen}
\definecolor{rossos}{rgb}{0.7,0,0.3}
\definecolor{violachiaro}{rgb}{1,0.6,1}
\definecolor{rossochiaro}{rgb}{1,0.6,0.6}
\definecolor{verdechiaro}{rgb}{0.6,1,0.6}
\definecolor{giallochiaro}{rgb}{1,1,0.6}
\definecolor{bluscuro}{rgb}{0.15, 0.2, 0.9}
\definecolor{verdes}{rgb}{0.1, 0.5, 0.1}
\definecolor{gold}{rgb}{1,0.84,0}
\definecolor{forestgreen}{rgb}{0.13,0.55,0.13}

\usepackage{lmodern}

\makeatletter
\ifcase \@ptsize \relax
  \newcommand{\miniscule}{\@setfontsize\miniscule{4}{5}}
\or
  \newcommand{\miniscule}{\@setfontsize\miniscule{5}{6}}
\or
  \newcommand{\miniscule}{\@setfontsize\miniscule{5}{6}}
\fi
\makeatother

\usepackage{lipsum}

\usepackage{bbold}

\newcommand{\nn}{\nonumber}

\newcommand{\be}{\begin{equation}}
\newcommand{\ee}{\end{equation}}
\newcommand{\bea}{\begin{eqnarray}}
\newcommand{\eea}{\end{eqnarray}}
\newcommand{\bc}{\begin{center}}
\newcommand{\ec}{\end{center}}

\usepackage{adjustbox}
\usepackage{array}
\usepackage{booktabs}
\usepackage{multirow}

\newcolumntype{R}[2]{%
    >{\adjustbox{angle=#1,lap=\width-(#2)}\bgroup}%
    l%
    <{\egroup}%
}

\title{\boldmath 
Inflation without gauge redundancy
}

\author[a,b]{Alfredo Urbano}

\affiliation[a]{I.N.F.N., Istituto Nazionale di Fisica Nucleare, sezione di Trieste\\SISSA, via Bonomea 265, I-34132 Trieste, Italy.}

\affiliation[b]{I.F.P.U., Institute  for  Fundamental Physics  of  the  Universe\\Via  Beirut  2, I-34014 Trieste, Italy.}

\emailAdd{alfredo.urbano@sissa.it}

\abstract{\\
In the context of gauge theories, observable quantities, if properly defined and computed, do not depend on the gauge-fixing procedure.
In this paper, we develop a formalism that implements this (apparent) tautology in the case of inflation.
As a simple application, we discuss Coleman-Weinberg ``hilltop'' inflation.
}

\begin{document}

\maketitle
\flushbottom

\section{Prolegomena}
\label{sec:intro}

In the context of gauge theories, observable quantities, {\it if properly defined and computed}, do not depend on the gauge-fixing procedure. 
The way in which this apparent tautology is concretely realized, however, is not always transparent.
We discuss in this paper the case of inflation. Consider a scalar filed $\phi$ with self-interacting potential $V(\phi)$ coupled to Einstein-Hilbert gravity by means of the curved-space classical action ($\bar{M}_{\rm Pl}$ is the reduced Planck mass,  $\bar{M}_{\rm Pl}^2 = 1/8\pi G_N$ with $G_N$ the Newton's constant, and $\mathcal{R}(g)$ the Ricci scalar associated with the metric $g$)
\begin{align}\label{eq:ExAction}
\mathcal{S}[\phi] = \int d^4 x\sqrt{-g}\left[-\frac{\bar{M}^2_{\rm Pl}}{2}\mathcal{R}(g) + 
\frac{1}{2}g^{\mu\nu}(\partial_{\mu}\phi)(\partial_{\nu}\phi) - V(\phi)
\right]\,,
\end{align}
with the homogeneous and isotropic (flat) Friedmann-Lema\^{\i}tre-Robertson-Walker (FLRW) metric 
$ds^2 = g_{\mu\nu}dx^{\mu}dx^{\nu} = dt^2 - a(t)^2(dr^2 + r^2d\Omega^2)$. 
Assuming the potential flat enough to guarantee the application of the slow-roll (SR) approximation, one can immediately make contact with  cosmological observables by introducing a handful of relevant parameters with a crystal-clear physical interpretation.
To be explicit, consider for instance the computation of the spectrum of primordial perturbations at CMB scales.
Perturbations  produced  by  generic  single-field  slow-roll models of inflation are typically well approximated by the following parametrization 
of the adiabatic scalar (subscript $_s$) and tensor (subscript $_t$) components\,\cite{Akrami:2018odb}
\begin{equation}
\mathcal{P}_{\mathcal{R}}(k) = A_s\left(\frac{k}{k_*}\right)^{n_s - 1 +\frac{\alpha}{2}\log\frac{k}{k_*}+\dots}\,,~~~~~~
\mathcal{P}_{t}(k) =A_t\left(\frac{k}{k_*}\right)^{n_t+\dots}\,,
\end{equation}
where the amplitudes $A_{s,t}$, the scalar and tensor spectral indices $n_{s,t}$ and the running parameter $\alpha \equiv dn_s/d\log k$  are implicitly evaluated at the pivot scale $k_*$. At leading order in the slow-roll expansion, these parameters are simple functions of the inflationary potential and its field derivatives
\begin{equation}
n_s \simeq 1+2\eta_V - 6\epsilon_V\,,~~~~~~~\alpha = -2\tau_V + 16\epsilon_V\eta_V - 24\epsilon_V^2\,,~~~~~~~n_t \simeq -\frac{r}{8}\,,
\end{equation}
and 
\begin{equation}
A_s = \mathcal{P}_{\mathcal{R}}(k_*) \simeq \frac{V}{24\pi^2 \epsilon_V \bar{M}_{\rm Pl}^4}\,,~~~~~~~A_t \simeq r A_s \simeq 16\epsilon_V A_s\,,
\end{equation}
where, as customary, we used the potential slow-roll parameters
\begin{equation}\label{eq:PotentialSRpar}
\epsilon_V = \frac{\bar{M}_{\rm Pl}^2}{2}\left(\frac{V^{\prime}}{V}\right)^2\,,~~~~~~~\eta_V = \bar{M}_{\rm Pl}^2\frac{V^{\prime\prime}}{V}
\,,~~~~~~~\tau_V = \bar{M}_{\rm Pl}^4\frac{V^{\prime}V^{\prime\prime\prime}}{V^2}\,,
\end{equation}
with $^{\prime} \equiv d/d\phi$. 
This astonishingly simple picture allows a direct comparison between experimental bounds on/measurements of cosmological observables  at CMB scales and any given inflationary
 model that fulfills the slow-roll approximation. 
 In addition to the self-interactions encoded in the potential, it is plausible to consider  scenarii in which the inflaton field also interacts with additional matter fields.
This is, actually, a mandatory requirement for realistic models in which one aims describing not only the accelerated expansion of the early Universe but also its post-inflationary reheating phase.  
In full generality, therefore, the tree-level inflaton potential is subject to quantum corrections due to matter loops.
In these cases it might be important to important to consider, instead of the tree-level potential, the quantum 
effective potential $V_{\rm eff}$ which embodies
 a sum over all of the 
one-particle irreducible (1PI) diagrams in the quantum theory\,\cite{Jackiw:1974cv}. 
 There are indeed many examples in the literature in which this procedure -- with the effective potential, if necessary, improved by means 
 of the renormalization group equations (RGEs) -- turns out to be the right way to go, and where radiative corrections play an important role to assess the validity of the corresponding inflationary model.
 A prominent example is that of Higgs inflation in which the Standard Model (SM) Higgs boson  can be successfully identified with the inflaton in the presence of a non-minimal coupling to Einstein-Hilbert gravity\,\cite{CervantesCota:1995tz,Bezrukov:2007ep}.  
Furthermore, quantum corrections to the inflaton potential could leave  peculiar imprints at scales much smaller than those probed by CMB observations. 
For instance, quantum corrections could be responsible for the presence of an approximate stationary inflection point before the end of inflation. 
The latter, in turn, could generate a peak in the scalar power spectrum at small scales.
In such cases, it is well-known that the slow-roll approximation does not capture all the relevant physics, and one is forced to solve the inflationary dynamics exactly\,\cite{Motohashi:2017kbs,Germani:2017bcs,Ballesteros:2017fsr}. 
Na\"{\i}vely, assuming the slow-roll approximation to be applicable, one is tempted to 
carry on the same analysis discussed before by simply replacing the tree-level potential in eq.\,(\ref{eq:PotentialSRpar}) with the effective potential $V_{\rm eff}$.
If the slow-roll approximation is not applicable, the na\"{\i}ve approach is to solve the equation of motion (EoM) of the inflaton (together with the Einstein field equations) by using the effective potential $V_{\rm eff}$ in eq.\,(\ref{eq:ExAction}) instead of the tree-level one.
 
In the context of a gauge theory, however, the effective potential is a gauge-dependent quantity\,\cite{Jackiw:1974cv} (see also refs.\,\cite{Andreassen:2014eha,Andreassen:2014gha} for a recent critical discussion).
Consequently, the very same gauge-dependence would be inherited by the slow-roll parameters in eq.\,(\ref{eq:PotentialSRpar}) 
 or by the solution of the EoM, thus polluting the comparison with observables.
 
From the above considerations, it appears clear that, at least at the conceptual level, when quantum corrections are important
 there must exist a way of discussing  inflationary dynamics -- both in the slow-roll limit and in its exact formulation -- 
 in which no spurious trace of gauge-dependence is left when observable quantities are computed.

\section{Inflationary dynamics without gauge redundancy}
\label{sec:Guage}

The Nielsen identity describes the gauge-fixing dependence of the (quantum) effective action $\mathcal{S}_{\rm eff}$, 
and represents the most important tool in discussing how to obtain gauge-independent quantities. 
In the presence of a single background field $\phi$, it reads\,\cite{Nielsen:1975fs,Fukuda:1975di,Aitchison:1983ns} 
\begin{equation}\label{eq:Nielsen}
\xi\frac{\partial\mathcal{S}_{\rm eff}[\phi,\xi]}{\partial\xi} = \int d^4 y\,K[\phi(y),\xi]\frac{\delta\mathcal{S}_{\rm eff}[\phi,\xi]}{\delta\phi(y)}\,,
\end{equation}
where $K[\phi(y),\xi]$ is a known functional of $\phi$ (see section\,\ref{sec:EffPot} for the corresponding expression in the context of an explicit model). 
Let us start our discussion with some general remarks. In this first part, we shall follow the line of reasoning of ref.\,\cite{Espinosa:2016nld}.
The starting point is the gauge invariance of the effective action, which we write in the form 
 \begin{align}\label{eq:Key}
 \xi\frac{dS_{\rm eff}[\phi,\xi]}{d\xi} = 
 \xi \frac{\partial S_{\rm eff}[\phi,\xi]}{\partial\xi} + 
 \int d^4y \frac{\delta S_{\rm eff}[\phi,\xi]}{\delta\phi(y)}\xi\frac{d\phi}{d\xi}
 \overset{!}{=} 0~~~~~\Longrightarrow~~~~~ \xi\frac{d}{d\xi}\phi(x,\xi) = -K[\phi(x,\xi),\xi]\,,
 \end{align}
 where the last condition follows from the Nielsen identity in eq.\,(\ref{eq:Nielsen}). 
 This condition is valid for {\it any} field configuration $\phi(x,\xi)$. 
 Eq.\,(\ref{eq:Key}) is telling us that the explicit $\xi$-dependence of the effective action (in general we have $\partial S_{\rm eff}/\partial\xi \neq 0$) is compensated by the 
 $\xi$-dependence  
 of the field configuration, $d\phi/d\xi \neq 0$, thus leading to a gauge-invariant result. 
 Eq.\,(\ref{eq:Key}) is the crucial ingredient that is needed to construct gauge-invariant observables. 
 It is worth emphasizing once again that eq.\,(\ref{eq:Key}) is not an extra condition that must be imposed on the dynamics;
 on the contrary, the $\xi$-dependence of any field configuration $\phi(x,\xi)$ does satisfy  eq.\,(\ref{eq:Key}) as a consequence of the gauge invariance of the effective action. 
 
If we take the functional derivative of eq.\,(\ref{eq:Nielsen}) with respect to $\phi(x)$ we find
\begin{equation}\label{eq:DerNielsen}
\xi\frac{\partial}{\partial\xi}\frac{\delta\mathcal{S}_{\rm eff}[\phi,\xi]}{\delta\phi(x)} = 
\int d^4 y\left\{
\frac{\delta^2\mathcal{S}_{\rm eff}[\phi,\xi]}{\delta\phi(y)\delta\phi(x)}\,K[\phi(y),\xi] +
\frac{\delta\mathcal{S}_{\rm eff}[\phi,\xi]}{\delta\phi(y)}\,\frac{\delta K[\phi(y),\xi]}{\delta\phi(x)}
\right\}\,.
\end{equation}
The second term on the right-hand side vanishes evaluated on a field configuration $\bar{\phi}(x,\xi)$ that solves the EoM $\delta\mathcal{S}_{\rm eff}[\phi,\xi]/\delta\phi = 0$. Notice that in general $\bar{\phi}(x,\xi)$ depends on $\xi$, as discussed before. 
Eq.\,(\ref{eq:DerNielsen}) becomes
\begin{equation}\label{eq:DerNielsen2}
\xi\frac{\partial}{\partial\xi}\left.\frac{\delta\mathcal{S}_{\rm eff}[\phi,\xi]}{\delta\phi(x)}\right|_{\phi = \bar{\phi}(x,\xi)} - 
\int d^4 y
\left.\frac{\delta^2\mathcal{S}_{\rm eff}[\phi,\xi]}{\delta\phi(y)\delta\phi(x)}\right|_{\phi = \bar{\phi}(x,\xi)}\,K[\bar{\phi}(y,\xi),\xi] = 0\,. 
\end{equation}
We now recognize that eq.\,(\ref{eq:DerNielsen2}) is nothing but the total derivative 
\begin{equation}\label{eq:EoMgauge}
\left.\xi\frac{d}{d\xi}\frac{\delta\mathcal{S}_{\rm eff}[\phi,\xi]}{\delta\phi(x)}\right|_{\phi = \bar{\phi}(x,\xi)} = 0\,,
\end{equation}
with the $\xi$-dependence of $\bar{\phi}(x,\xi)$ that satisfies the equation
\begin{equation}\label{eq:FieldResc}
\xi\frac{d}{d\xi}\bar{\phi}(x,\xi) = -K[\bar{\phi}(x,\xi),\xi]\,,
\end{equation}
as derived in eq.\,(\ref{eq:Key}).
Eq.\,(\ref{eq:FieldResc}) describes how a solution of the EoM changes when varying the gauge-fixing parameter $\xi$.
In words, if we denote with $\bar{\phi}(x,\xi)$ a solution of the EoM for a given value of $\xi$, the solution that corresponds to the
 shifted value of the gauge-fixing parameter $\xi \to \xi +d\xi$ will be $\bar{\phi}(x,\xi) + d\bar{\phi}(x,\xi)$
 with $d\bar{\phi}(x,\xi) = -K[\bar{\phi}(x,\xi),\xi]d\log\xi$.
The fact that the total $\xi$-derivative in eq.\,(\ref{eq:EoMgauge}) is zero means that the explicit $\xi$-dependence of the EoM is compensated by 
the $\xi$-dependence of its solution -- the latter satisfying  eq.\,(\ref{eq:FieldResc}) -- thus leading to a  gauge-invariant result.

The same reasoning can be applied to the energy momentum tensor
\begin{equation}\label{eq:EM}
T^{\mu\nu}(x) = -\frac{2}{\sqrt{-g}}\frac{\delta\mathcal{S}_{\rm eff}[\phi,\xi]}{\delta g_{\mu\nu}(x)}\,,
\end{equation}
as discussed in ref.\,\cite{Espinosa:2016nld}.
By taking the variation of eq.\,(\ref{eq:Nielsen}) with respect to $g_{\mu\nu}$, we find
\begin{equation}
\xi\frac{\partial}{\partial\xi}T^{\mu\nu}(x) = -\frac{2}{\sqrt{-g}}
\int d^4 y\left\{
\frac{\delta^2\mathcal{S}_{\rm eff}[\phi,\xi]}{\delta\phi(y)\delta g_{\mu\nu}(x)}\,K[\phi(y),\xi] +
\frac{\delta\mathcal{S}_{\rm eff}[\phi,\xi]}{\delta\phi(y)}\,\frac{\delta K[\phi(y),\xi]}{\delta g_{\mu\nu}(x)}
\right\}\,.
\end{equation}
Evaluated on a solution of the EoM $\bar{\phi}(x,\xi)$ such that eq.\,(\ref{eq:FieldResc}) is satisfied, we find
\begin{equation}\label{eq:EMgauge}
\left.\xi\frac{d}{d\xi}T^{\mu\nu}(x)\right|_{\phi = \bar{\phi}(x,\xi)} = 0\,.
\end{equation}
As before, the explicit gauge dependence is compensated by the change of the field value as $\xi$ changes. 


It is instructive to see how eqs.\,(\ref{eq:EoMgauge},\,\ref{eq:EMgauge}) explicitly work in the context of a concrete example.
To this end, we consider the following effective action
\begin{equation}\label{eq:QuadraticS}
\mathcal{S}_{\rm eff}[\phi,\xi] = \int d^4 x\sqrt{-g}\left[
- V_{\rm eff} + \frac{Z_{\rm eff}}{2}g^{\mu\nu}(\partial_{\mu}\phi)(\partial_{\nu}\phi)  + \dots
\right]\,,
\end{equation}
where we only included terms up to the second order in the derivative expansion.
In full generality, 
we have $Z_{\rm eff} \equiv Z_{\rm eff}(\phi,\{\lambda_i,\textsl{g}_i\},\xi,\mu)$ and $V_{\rm eff} \equiv V_{\rm eff}(\phi,\{\lambda_i,\textsl{g}_i\},\xi,\mu)$, meaning that
both 
$Z_{\rm eff}$ and $V_{\rm eff}$ depend on the field $\phi$, the gauge-fixing parameter $\xi$ and a set of dimensionless couplings $\{\lambda_i,\textsl{g}_i\}$.
The latter are present because $Z_{\rm eff}$ and $V_{\rm eff}$ are generated by quanta of the fields coupled to $\phi$ running in the loops.
Furthermore, $Z_{\rm eff}$ and $V_{\rm eff}$ depend on the renormalization scale $\mu$. This is because the quantum corrections entering in the computation of the effective action are plagued by ultraviolet (UV) infinities that must be renormalized, as customary, once a regularization procedure and a subtraction scheme are chosen.
Furthermore, $Z_{\rm eff}$ and $V_{\rm eff}$ admit a loop expansion in powers of $\hbar$ of the form
\begin{equation}
Z_{\rm eff} = 1 + \underbrace{Z^{(1)}_{\rm eff}(\phi,\{\lambda_i,\textsl{g}_i\},\xi,\mu)}_{\rm 1\,loop\,kinetic\,correction} + \dots\,,~~~~~~~~~~
V_{\rm eff} = \underbrace{V_0(\phi,\{\lambda_i\})}_{\rm tree\,level} + 
\underbrace{V^{(1)}_{\rm eff}(\phi,\{\lambda_i,\textsl{g}_i\},\xi,\mu)}_{\rm 1\,loop\,Coleman\,Weinberg} + \dots\,.
\end{equation}
As far as the expansion of $Z_{\rm eff}$ is concerned, we are assuming here a canonically normalized kinetic term such that $Z_{\rm eff}=1$ at the tree level.
We shall compute $Z_{\rm eff}$ and $V_{\rm eff}$ at one loop in the context of an explicit model in section\,\ref{sec:CHI}.
Finally, it is also possible improve the $L$-loop effective action by evolving all dimensionless couplings by means of the ($L$+1)-loop $\beta$ functions
(thus obtaining an expression that resums all $L^{\rm th}$-to-leading logarithms\,\cite{Ford:1992mv}).
For the moment, we only need the generic expression for the quadratic effective action in eq.\,(\ref{eq:QuadraticS}).
In addition, we take the derivative expansion of eq.\,(\ref{eq:FieldResc})
\begin{equation}\label{eq:FieldRescExp}
-\xi\frac{d}{d\xi}\bar{\phi}(x,\xi) = C(\bar{\phi},\xi) + D(\bar{\phi},\xi)g^{\mu\nu}(\partial_{\mu}\phi)(\partial_{\nu}\phi) +
\frac{\tilde{D}(\bar{\phi},\xi)}{\sqrt{-g}}\partial_{\nu}\left[
\sqrt{-g}g^{\mu\nu}(\partial_{\mu}\phi) 
\right] + \dots\,.
\end{equation}
Notice that we are considering a generic (non-dynamical) background metric $g_{\mu\nu}(x)$.
Inserting eq.\,(\ref{eq:QuadraticS}) and the expansion for $K$ on the right-hand side of eq.\,(\ref{eq:FieldRescExp}) into eq.\,(\ref{eq:Nielsen}), we can
compare the two sides of the Nielsen identity and match terms with zero and two derivatives\,\cite{Metaxas:1995ab,Garny:2012cg}. 
We find (omitting functional dependencies for the sake of compactness, and indicating with $^{\prime}$ the derivative with respect to $\phi$) 
\begin{align}
\xi\frac{\partial V_{\rm eff}}{\partial\xi} &= CV_{\rm eff}^{\prime}\,,\label{eq:ZerothNiel}\\
\xi\frac{\partial Z_{\rm eff}}{\partial\xi} &= CZ_{\rm eff}^{\prime} + 2C^{\prime}Z_{\rm eff} - 2DV_{\rm eff}^{\prime} + 2(\tilde{D}V_{\rm eff}^{\prime})^{\prime}\,,\label{eq:FirstNiel}\\
\xi\frac{\partial V^{\prime}_{\rm eff}}{\partial\xi} &= C^{\prime}V_{\rm eff}^{\prime}+CV_{\rm eff}^{\prime\prime}\,,\\
\xi\frac{\partial Z^{\prime}_{\rm eff}}{\partial\xi} &=3C^{\prime}Z_{\rm eff}^{\prime} +
CZ_{\rm eff}^{\prime\prime} + 2C^{\prime\prime}Z_{\rm eff} -2D^{\prime}V_{\rm eff}^{\prime}-
2DV_{\rm eff}^{\prime\prime}+2(\tilde{D}V_{\rm eff}^{\prime})^{\prime\prime}\,.
\end{align}
In the following, we shall also need  space-time derivative of eq.\,(\ref{eq:FieldRescExp}). We find, up to terms with at most 
two derivatives
\begin{align}
-\xi\frac{d}{d\xi}(\partial_{\mu}\bar{\phi}) &= C^{\prime}(\partial_{\mu}\bar{\phi})\,,\label{eq:FieldRescExp2}\\
-\xi\frac{d}{d\xi}(\partial_{\mu}\partial_{\nu}\bar{\phi}) &= C^{\prime\prime}
(\partial_{\mu}\bar{\phi})(\partial_{\nu}\bar{\phi}) + C^{\prime}(\partial_{\mu}\partial_{\nu}\bar{\phi})\,.\label{eq:FieldRescExp3}
\end{align}
The EoM is
\begin{align}
\frac{\delta\mathcal{S}_{\rm eff}}{\delta\phi} = \frac{Z_{\rm eff}^{\prime}}{2}(\partial_{\mu}\phi)(\partial^{\mu}\phi) + \frac{Z_{\rm eff}}{\sqrt{-g}}
\partial_{\nu}\left[\sqrt{-g}g^{\mu\nu}\left(\partial_{\mu}\phi\right)\right] + V_{\rm eff}^{\prime}\,,\label{eq:MasterEoM}
\end{align} 
and we can now check the validity of eq.\,(\ref{eq:EoMgauge}). For illustrative purposes, we shall do this computation explicitly. We write
\begin{align}
\xi\frac{d}{d\xi}\frac{\delta\mathcal{S}_{\rm eff}}{\delta\phi} = & +
\xi\frac{\partial}{\partial\xi}\left(\frac{\delta\mathcal{S}_{\rm eff}}{\delta\phi}\right) +  
\frac{\partial}{\partial\phi}\left(\frac{\delta\mathcal{S}_{\rm eff}}{\delta\phi}\right)\xi \frac{d\phi}{d\xi} + 
\frac{\partial}{\partial(\partial_{\mu}\phi)}\left(\frac{\delta\mathcal{S}_{\rm eff}}{\delta\phi}\right)\xi \frac{d}{d\xi}(\partial_{\mu}\phi)\label{eq:EoMCheck}\\&
+ 
\frac{\partial}{\partial(\partial_{\nu}\partial_{\mu}\phi)}\left(\frac{\delta\mathcal{S}_{\rm eff}}{\delta\phi}\right)\xi \frac{d}{d\xi}(\partial_{\nu}\partial_{\mu}\phi) =\nn\\
= & +\frac{1}{2}\left(\xi\frac{\partial Z_{\rm eff}^{\prime}}{\partial\xi} + Z_{\rm eff}^{\prime\prime}\xi\frac{d\phi}{d\xi}\right)(\partial_{\mu}\phi)(\partial^{\mu}\phi) 
+\left(\xi\frac{\partial Z_{\rm eff}}{\partial\xi} + Z_{\rm eff}^{\prime}\xi\frac{d\phi}{d\xi}\right)\frac{1}{\sqrt{-g}}
\partial_{\nu}\left[\sqrt{-g}g^{\mu\nu}\left(\partial_{\mu}\phi\right)\right]\nn \\
& + \left[Z_{\rm eff}^{\prime}(\partial^{\mu}\phi) + \frac{Z_{\rm eff}}{\sqrt{-g}}\partial_{\nu}\left(\sqrt{-g}g^{\mu\nu}\right)
\right]\xi\frac{d}{d\xi}(\partial_{\mu}\phi) + Z_{\rm eff}g^{\mu\nu}\xi\frac{d}{d\xi}(\partial_{\nu}\partial_{\mu}\phi) + \xi\frac{\partial V_{\rm eff}^{\prime}}{\partial\xi} +  V_{\rm eff}^{\prime\prime}\xi\frac{d\phi}{d\xi}\nn\,.
\end{align}
It is important to keep in mind that we are working at the quadratic order in the field derivatives. 
This implies important simplifications in the previous expression.
Take for instance the term $\xi(d\phi/d\xi)(\partial_{\mu}\phi)(\partial^{\mu}\phi)$ in the first line of eq.\,(\ref{eq:EoMCheck}); since this is already $O(\partial^2)$, 
we can use   eq.\,(\ref{eq:FieldRescExp}) at the zeroth order, namely $\xi(d\phi/d\xi) = - C$. Similar arguments work for the other terms, and  
it is easy to see that one never uses the coefficients $D$ and $\tilde{D}$ 
in eq.\,(\ref{eq:FieldRescExp}) since they always contribute to higher derivative orders.
We have
\begin{align}
\xi\frac{d}{d\xi}\frac{\delta\mathcal{S}_{\rm eff}}{\delta\phi}
 = & +
\left(
\frac{\xi}{2}\frac{\partial Z_{\rm eff}^{\prime}}{\partial\xi} - C\frac{Z_{\rm eff}^{\prime\prime}}{2} - C^{\prime}Z- C^{\prime\prime}Z
\right)
(\partial_{\mu}\phi)(\partial^{\mu}\phi) \nn\\
& + \left(
\xi\frac{\partial Z_{\rm eff}}{\partial\xi} - CZ_{\rm eff}^{\prime} - C^{\prime}Z
\right)\frac{1}{\sqrt{-g}}\partial_{\nu}\left[
\sqrt{-g}g^{\mu\nu}(\partial_{\mu}\phi)
\right] + C^{\prime}V^{\prime}_{\rm eff}\,.
\end{align}
We can now use the fact that in eq.\,(\ref{eq:EoMgauge}) the total derivative is evaluated for the solution of the EoM $\phi = \bar{\phi}$.
It means that we can use $\xi(\partial Z_{\rm eff}/\partial\xi) = CZ_{\rm eff}^{\prime} + 2C^{\prime}Z_{\rm eff}$ and  $\xi(\partial Z^{\prime}_{\rm eff}/\partial\xi) =3C^{\prime}Z_{\rm eff}^{\prime} +
CZ_{\rm eff}^{\prime\prime} + 2C^{\prime\prime}Z_{\rm eff}$.
This is because and 
in eq.\,(\ref{eq:FirstNiel}) $V^{\prime}_{\rm eff}$, as a consequence of the EoM in eq.\,(\ref{eq:MasterEoM}), actually counts 
as a second-order derivative term thus contributing at higher order in the derivative expansion.
 We find
\begin{align}
\left.\xi\frac{d}{d\xi}\frac{\delta\mathcal{S}_{\rm eff}}{\delta\phi}\right|_{\phi = \bar{\phi}} = C^{\prime}\left\{
\frac{Z_{\rm eff}^{\prime}}{2}(\partial_{\mu}\bar{\phi})(\partial^{\mu}\bar{\phi}) + \frac{Z_{\rm eff}}{\sqrt{-g}}
\partial_{\nu}\left[\sqrt{-g}g^{\mu\nu}(\partial_{\mu}\bar{\phi})\right] + V_{\rm eff}^{\prime}\right\} = 0\,,
\end{align}
which is zero because $\bar{\phi}$ solves the EoM\,(\ref{eq:MasterEoM}). 
In very similar fashion one can check explicitly the gauge invariance of the energy-momentum tensor in eq.\,(\ref{eq:EMgauge}) by using the expression
\begin{align}\label{eq:ExplT}
T_{\mu\nu} = Z_{\rm eff}(\partial_{\mu}\phi)(\partial_{\nu}\phi) - g_{\mu\nu}\left[\frac{Z_{\rm eff}}{2}
(\partial_{\rho}\phi)(\partial^{\rho}\phi)  - V_{\rm eff}
\right]\,,~~~~~\left.\xi\frac{d}{d\xi}T^{\mu\nu}\right|_{\phi = \bar{\phi}} = 0\,.
\end{align}
Gauge invariance of the energy-momentum tensor is valid for each of its components. 

\subsection{The Hubble rate}\label{sec:HubbleGauge}

If we take the flat FLRW  metric $ds^2 = g_{\mu\nu}dx^{\mu}dx^{\nu} = dt^2 - a(t)^2(dr^2 + r^2d\Omega^2) \equiv dt^2 - a(t)^2\kappa_{ij}dx^{i}dx^j$ the 
Einstein field equations read
\begin{align}
\mathcal{G}_{\mu\nu} \equiv \mathcal{R}_{\mu\nu} - \frac{1}{2}g_{\mu\nu}\mathcal{R} = 8\pi G_N T_{\mu\nu}\,,~~~~{\rm with}~~~
\mathcal{G}_{\mu\nu} = 
\left\{
\begin{array}{cc}
\mathcal{G}_{00} =  & 3(\dot{a}/a)^2 = 3H^2  \\
\mathcal{G}_{ij} = &  -\kappa_{ij}(\dot{a}^2 + 2a\ddot{a})
\end{array}
\right.\label{eq:EinsteinFieldEq}
\end{align}
and gauge invariance of the Hubble rate $H(t)\equiv \dot{a}/a$ during inflation follows immediately  from gauge invariance of the  energy-momentum tensor. 
Interchangeably, we shall use the short-hand notation $\dot{} \equiv d/dt$.
At the quadratic order in the expansion of the effective action, Einstein field equations\,(\ref{eq:EinsteinFieldEq})  are
\begin{align}
3H^2 &= \frac{1}{\bar{M}_{\rm Pl}^2}\left(
\frac{Z_{\rm eff}}{2}\dot{\phi}^2 + V_{\rm eff}
\right)\,,\label{eq:FirstEin}\\
 -(\dot{a}^2 + 2a\ddot{a}) & = \frac{a^2}{\bar{M}_{\rm Pl}^2}
\left(
\frac{Z_{\rm eff}}{2}\dot{\phi}^2 -  V_{\rm eff}
\right)\,,\label{eq:SecondEin}
\end{align}
where now notice that we are considering only the homogeneous mode of the scalar field, $\partial_{\mu}\phi = (\dot{\phi},0)$ (inflation rapidly smooths out spatial variations, and the approximation is certainly good after few $e$-folds of accelerated expansion).
For later use, let us derive a number of simple relations.
If we trade the cosmic time variable $t$ for the number of $e$-folds $N_e$ via $dN_e = Hdt$, the first equation can be rewritten in the form
\begin{align}\label{eq:HJ}
3H^2\left[
1-\frac{Z_{\rm eff}}{6\bar{M}_{\rm Pl}^2}\left(\frac{d\phi}{dN_e}\right)^2
\right] = \frac{V_{\rm eff}}{\bar{M}_{\rm Pl}^2}\,.
\end{align}
From eqs.\,(\ref{eq:FirstEin},\ref{eq:SecondEin}) we obtain the time derivative of the Hubble rate. 
We find
\begin{equation}\label{eq:Hdot}
\dot{H} = -\frac{Z_{\rm eff}}{2\bar{M}_{\rm Pl}^2}\dot{\phi}^2\,.
\end{equation}
We can check the gauge independence of $\dot{H}$ in the usual way by computing
\begin{align}
\left.\xi\frac{d\dot{H}}{d\xi}\right|_{\phi = \bar{\phi}} = -\frac{1}{2\bar{M}_{\rm Pl}^2}\Bigg(
\underbrace{\xi\frac{\partial Z_{\rm eff}}{\partial\xi}\dot{\phi}^2}_{(CZ_{\rm eff}^{\prime} + 2C^{\prime}Z_{\rm eff})\dot{\phi}^2} + 
\underbrace{Z_{\rm eff}^{\prime}\xi\frac{d\phi}{d\xi}\dot{\phi}^2}_{-CZ_{\rm eff}^{\prime}\dot{\phi}^2} + 
\underbrace{2Z_{\rm eff}\dot{\phi}\xi\frac{d\dot{\phi}}{d\xi}}_{-2C^{\prime}Z_{\rm eff}\dot{\phi}^2}
\Bigg) = 0\,,
\end{align}
where again we limited the analysis at the second order in the derivative expansion.
Finally, we write the EoM\,(\ref{eq:MasterEoM}) in the form
\begin{align}
\frac{\delta\mathcal{S}_{\rm eff}}{\delta\phi} & = 
Z_{\rm eff}\ddot{\phi} + 3Z_{\rm eff}H\dot{\phi} + \frac{Z_{\rm eff}^{\prime}}{2}\dot{\phi}^2 + V_{\rm eff}^{\prime} = 0\,\label{eq:EoMt}\\
& = Z_{\rm eff}\frac{d^2\phi}{dN_e^2} + 3Z_{\rm eff}\frac{d\phi}{dN_e} - \frac{Z^2_{\rm eff}}{2\bar{M}_{\rm Pl}^2}
\left(\frac{d\phi}{dN_e}\right)^3
+\frac{Z_{\rm eff}^{\prime}}{2}\left(\frac{d\phi}{dN_e}\right)^2 + \left[
3\bar{M}_{\rm Pl}^2 - \frac{Z_{\rm eff}}{2}\left(\frac{d\phi}{dN_e}\right)^2 
\right]\frac{V^{\prime}_{\rm eff}}{V_{\rm eff}} = 0\,,\label{eq:EoMN}
\end{align}
with $\phi = \phi(t)$ or $\phi = \phi(N_e)$, respectively. Gauge invariance of the EoM in the homogeneous limit follows from the discussion in eq.\,(\ref{eq:EoMCheck}).
In the slow-roll approximation the EoM\,(\ref{eq:EoMt}) takes the simplified form 
\begin{align}\label{eq:EoMSR}
\frac{\delta\mathcal{S}_{\rm eff}}{\delta\phi}_{\rm SR} = 3Z_{\rm eff}H\dot{\phi} + V_{\rm eff}^{\prime} \overset{{\rm SR}}{=} 0\,,
\end{align}
where the symbol $\overset{{\rm SR}}{=}$ indicates that the equality is valid in the slow-roll approximation.
The time-time component of Einstein field equations, eq.\,(\ref{eq:FirstEin}), reads
\begin{equation}\label{eq:SRFried}
3H^2 \overset{{\rm SR}}{=} \frac{V_{\rm eff}}{\bar{M}_{\rm Pl}^2}\,.
\end{equation}
It is straightforward but useful to check that gauge invariance is preserved in the slow-roll limit. We find
\begin{align}
3\left.\xi\frac{dH^2}{d\xi}\right|_{\phi = \bar{\phi}} \overset{{\rm SR}}{=}  \frac{1}{\bar{M}_{\rm Pl}^2}\Bigg(
\underbrace{\xi\frac{\partial V_{\rm eff}}{\partial\xi}}_{CV_{\rm eff}^{\prime}} + \underbrace{V_{\rm eff}^{\prime}\xi \frac{d\phi}{d\xi}}_{-V_{\rm eff}^{\prime}C} \Bigg) = 0\,,
\end{align}
and
\begin{align}
\left.\xi\frac{d}{d\xi}\frac{\delta\mathcal{S}_{\rm eff}}{\delta\phi}_{\rm SR}\right|_{\phi = \bar{\phi}}& =\left.
3H\dot{\phi}\xi\frac{\partial Z_{\rm eff}}{\partial\xi} + 3Z_{\rm eff}^{\prime}H\dot{\phi}\xi\frac{d\phi}{d\xi} 
+3Z_{\rm eff}H\xi\frac{d\dot{\phi}}{d\xi} +\xi\frac{\partial V_{\rm eff}^{\prime}}{\partial\xi} + V_{\rm eff}^{\prime\prime}\xi\frac{d\phi}{d\xi}\right|_{\phi = \bar{\phi}}\\
&= 3H\frac{d\bar{\phi}}{dt}\Bigg(
\underbrace{\xi\frac{\partial Z_{\rm eff}}{\partial\xi}}_{CZ_{\rm eff}^{\prime} + 2Z_{\rm eff}C^{\prime}} - CZ_{\rm eff}^{\prime} - Z_{\rm eff}C^{\prime}
\Bigg) + C^{\prime}V_{\rm eff}^{\prime} = C^{\prime}\left(
3HZ_{\rm eff}\frac{d\bar{\phi}}{dt} + V_{\rm eff}^{\prime}
\right)  = 0\,,\nn
\end{align}
where the last term vanishes since $\bar{\phi}$ satisfies the EoM in the slow-roll limit, eq.\,(\ref{eq:EoMSR}).

\subsection{The Hubble parameters and their slow-roll limit}\label{sec:SlowRollPar}

The Hubble parameter $\epsilon_H$ is defined as 
\begin{align}\label{eq:HubbleEps}
\epsilon_H \equiv - \frac{\dot{H}}{H^2} = \frac{Z_{\rm eff}\dot{\phi}^2}{2\bar{M}_{\rm Pl}^2 H^2} = \frac{Z_{\rm eff}}{2\bar{M}_{\rm Pl}^2}\left(\frac{d\phi}{dN_e}\right)^2\,,
\end{align}
that is a gauge-invariant quantity as a consequence of the discussion in section\,\ref{sec:HubbleGauge}.
Eq.\,(\ref{eq:FirstEin}) and eq.\,(\ref{eq:HJ}) now read
\begin{align}
3H^2 &= \rho_{\phi}\,,\\
3H^2\left(
1-\frac{\epsilon_H}{3}
\right) &= \frac{V_{\rm eff}}{\bar{M}_{\rm Pl}^2}\,.\label{eq:EinstEps}
\end{align}
Eq.\,(\ref{eq:EinstEps}) is the  usual Hamilton-Jacobi equation for a canonical scalar field, and the dependence on the kinetic factor $Z_{\rm eff}$
 has been absorbed into the expression in eq.\,(\ref{eq:HubbleEps}) for the Hubble parameter $\epsilon_H$. 
 From eq.\,(\ref{eq:ExplT}) we can derive the equation of state of $\phi$.
 We find for energy density and pressure  $T_{00} = \rho_{\phi} = Z_{\rm eff}\dot{\phi}^2/2 + V_{\rm eff}$ and 
 $T^{i}_{~i}/3 = P_{\phi} = Z_{\rm eff}\dot{\phi}^2/2 - V_{\rm eff}$. Consequently, by means of eq.\,(\ref{eq:HubbleEps}) and eq.\,(\ref{eq:EinstEps}) 
 we find the relations
 \begin{align}
\rho_{\phi} + P_{\phi}  = Z_{\rm eff}\dot{\phi}^2 \,,~~~~~~~\rho_{\phi} + 3P_{\phi} = 2Z_{\rm eff}\dot{\phi}^2 - 2V_{\rm eff} = -6\bar{M}_{\rm Pl}^2 H^2(1-\epsilon_{H})\,,
 \end{align}
 and from eq.\,(\ref{eq:SecondEin}) the cosmic acceleration takes the form 
 \begin{align}
 \frac{\ddot{a}}{a} = H^2(1-\epsilon_{H})~~~~\Longrightarrow~~~~{\rm condition\,for\,accelerated\,expansion}\,~\epsilon_{H} < 1\,,
\end{align} 
 so that the condition for an inflating Universe ($\epsilon_{H} < 1$) remains the usual one. 
 In terms of $\epsilon_H$, the EoM as a function of the number of $e$-folds, eq.\,(\ref{eq:EoMN}), can be rewritten as
 \begin{align}
\frac{\delta\mathcal{S}_{\rm eff}}{\delta\phi} & = Z_{\rm eff}\frac{d^2\phi}{dN_e^2} + 
(3-\epsilon_H)Z_{\rm eff}\frac{d\phi}{dN_e} + \bar{M}_{\rm Pl}^2\epsilon_H\frac{Z_{\rm eff}^{\prime}}{Z_{\rm eff}} +
\bar{M}_{\rm Pl}^2(3-\epsilon_H)\frac{V^{\prime}_{\rm eff}}{V_{\rm eff}} = 0\,.
\end{align}
 We can now take the slow-roll approximation. The slow-roll limit of the Hubble parameter $\epsilon_{H}$ defines the potential slow-roll parameter $\epsilon_V$.
 We find, by means of eq.\,(\ref{eq:EoMSR}) and eq.\,(\ref{eq:SRFried})
 \begin{align}\label{eq:GaugeInvEpsilon}
 \epsilon_H = \frac{Z_{\rm eff}\dot{\phi}^2}{2\bar{M}_{\rm Pl}^2 H^2}  \overset{{\rm SR}}{=} \frac{\bar{M}_{\rm Pl}^2}{2Z_{\rm eff}}\left(
 \frac{V_{\rm eff}^{\prime}}{V_{\rm eff}}
 \right)^2\equiv \epsilon_V\,.
 \end{align}
 As usual, we can check gauge invariance of $\epsilon_V$ by computing $\left.\xi d\epsilon_V/d\xi\right|_{\phi = \bar{\phi}} = 0$. The computation is analogue to what already discussed.
Contrary to the potential slow-roll parameter  $\epsilon_V$ introduced in eq.\,(\ref{eq:PotentialSRpar}), the one defined in eq.\,(\ref{eq:GaugeInvEpsilon}) is, therefore, gauge invariant while maintaining the same physical meaning.

We now move to consider the Hubble parameter $\eta_H$. 
We have
\begin{align}\label{eq:EtaGaugeInv}
\eta_H & = -\frac{\ddot{H}}{2H\dot{H}} = \epsilon_H - \frac{1}{2}\frac{d\log \epsilon_H}{dN} = \epsilon_H - \frac{1}{2H}\frac{d\log \epsilon_H}{dt} 
 = -\frac{1}{HZ_{\rm eff}\dot{\phi}}\left(
\frac{Z_{\rm eff}^{\prime}\dot{\phi}^2}{2} + Z_{\rm eff}\ddot{\phi}
\right)\nn\\
& = 3 + \frac{1}{2Z_{\rm eff}(d\phi/dN_e)}\left[
6\bar{M}_{\rm Pl}^2 - Z_{\rm eff}\left(\frac{d\phi}{dN_e}\right)^2
\right]\frac{V_{\rm eff}^{\prime}}{V_{\rm eff}} 
= 3 + \frac{\bar{M}_{\rm Pl}^2\left(3-\epsilon_H\right)}{Z_{\rm eff}(d\phi/dN_e)}\frac{V_{\rm eff}^{\prime}}{V_{\rm eff}}\,,
\end{align}
where on the right-hand side we used eq.\,(\ref{eq:Hdot}) together with  
\begin{align}
\ddot{H} = -\frac{\dot{\phi}}{2\bar{M}_{\rm Pl}^2}\left(
Z_{\rm eff}^{\prime}\dot{\phi}^2 + 2Z_{\rm eff}\ddot{\phi}
\right)\,.
\end{align}
Clearly, the parameter $\eta_H$ defined in eq.\,(\ref{eq:EtaGaugeInv}) is gauge invariant. 
The slow-roll limit of the combination $\epsilon_H + \eta_H$ defines the potential slow-roll parameter $\eta_V$.  
In formulas, we find
\begin{align}\label{eq:GaugeInvEta}
\epsilon_H + \eta_H \overset{{\rm SR}}{=} \frac{\bar{M}_{\rm Pl}^2}{Z_{\rm eff}}\left(
\frac{V_{\rm eff}^{\prime\prime}}{V_{\rm eff}} - \frac{Z_{\rm eff}^{\prime}V_{\rm eff}^{\prime}}{2Z_{\rm eff}V_{\rm eff}}
\right)\equiv \eta_V\,.
\end{align}
For completeness, it is again possible to check explicitly the gauge independence of $\eta_V$, $\left.\xi(d\eta_V/d\xi)\right|_{\phi = \bar{\phi}}$.\footnote{For this check, one needs
\begin{align}
\xi\frac{\partial V_{\rm eff}^{\prime\prime}}{\partial\xi} = C^{\prime\prime}V_{\rm eff}^{\prime} + 2C^{\prime}V_{\rm eff}^{\prime\prime} + CV_{\rm eff}^{\prime\prime\prime}\,.
\end{align}}
In the following we shall use, instead of the potential slow-roll parameter  $\eta_V$ introduced in eq.\,(\ref{eq:PotentialSRpar}), its gauge-invariant version in eq.\,(\ref{eq:GaugeInvEta}).

\subsection{The number of $e$-folds}\label{sec:Ne}

Gauge invariance of the number of $e$-folds is somehow a bit more tricky. 
We start from the usual definition
\begin{align}\label{eq:NofEfolds}
N_e \equiv \int_{t_{\rm in}}^{t_{\rm end}}H dt = \int_{\phi_{\rm in}}^{\phi_{\rm end}}\frac{H}{\dot{\phi}}d\phi\,,
\end{align}
where the integration extends between the beginning (subscript $_{\rm in}$) and the end (subscript $_{\rm end}$) of inflation, and where 
in the second step we changed variables in the field space by means of $dt = (1/\dot{\phi})d\phi$. 
Consider eq.\,(\ref{eq:NofEfolds}) in its differential form, $dN_e = (H/\dot{\phi})d\phi$.
Under an infinitesimal change of the gauge-fixing parameter $\xi \to \xi + d\xi$, we have $\phi \to \phi - Kd\log\xi$.  
Consequently, the measure $d\phi$ changes according to $d\phi \to d\phi - (K^{\prime}d\log\xi)d\phi$, and $dN_e$ is left invariant since 
\begin{align}
dN_e \overset{\xi \to \xi + d\xi }{\to} \frac{H}{\dot{\phi}}d\phi + \left(\frac{H}{\dot{\phi}}K^{\prime}d\log\xi\right)d\phi - \left(\frac{H}{\dot{\phi}}K^{\prime}d\log\xi\right)d\phi = dN_e\,.
\end{align}
This implies that the integrated quantity in eq.\,(\ref{eq:NofEfolds}) is gauge-fixing invariant provided that the integration limits  transform according to eq.\,(\ref{eq:Key}). 
If we take the slow-roll approximation, we find
\begin{align}
N_e =  \int_{\phi_{\rm in}}^{\phi_{\rm end}}\frac{H}{\dot{\phi}}d\phi \overset{{\rm SR}}{=} - \frac{1}{\bar{M}_{\rm Pl}^2}
\int_{\phi_{\rm in}(\xi)}^{\phi_{\rm end}(\xi)}\left(
\frac{Z_{\rm eff}V_{\rm eff}}{V^{\prime}_{\rm eff}}
\right)d\phi\,,
\end{align}
which generalizes in a gauge-fixing invariant way the standard definition. 

\subsection{The scalar power spectrum and the Mukhanov-Sasaki equation}

In order to study the primordial power spectrum we need to add spatially-dependent perturbations  to the homogenous classical inflaton trajectory (here indicated with an extra subscript $_0$), $\phi(t,\mathbf{x}) = \phi_0(t) +\delta\phi(t,\mathbf{x})$.
As a function of the cosmic time $t$, $\phi_0(t)$ solves the EoM\,(\ref{eq:EoMt}).
We also include metric perturbations.
We write the metric of the perturbed Universe in the longitudinal  gauge  (a.k.a. conformal newtonian gauge)\footnote{Of course, here ``gauge'' refers to 
invariance of gravity under generic coordinate transformations from a local reference frame to another.}
\begin{align}\label{eq:PerturbedMetric}
ds^2 = 
[g_{\mu\nu}^{(0)} + \delta g_{\mu\nu}]dx^{\mu}dx^{\nu} = 
 (1+2\Phi)dt^2 - (1-2\Phi)a(t)^2\kappa_{ij}dx^i dx^j\,,
\end{align}
where $g_{\mu\nu}^{(0)}$ is the unperturbed metric (namely  the FLRW metric introduced in section\,\ref{sec:HubbleGauge}) and $\Phi = \Phi(t,\mathbf{x})$ is the newtonian gravitational potential.\footnote{
In the longitudinal gauge we have two degrees of freedom in the scalar perturbations of the metric. However, since the energy-momentum tensor does not have any non-diagonal spatial component (no stress), the 
two degrees of freedom are constrained by the off-diagonal Einstein field equations, leaving behind only the gravitational potential as the only degree of freedom in  eq.\,(\ref{eq:PerturbedMetric}).
}
Our goal is to find a gauge-fixing invariant expression for  the power spectrum of the gauge-invariant comoving curvature perturbation
\begin{equation}
\mathcal{R} \equiv \Phi + H\frac{\delta\phi}{\dot{\phi}_0}\,.
\end{equation} 
We start from the perturbed Einstein field equations.
The perturbations $\delta\phi$ and $\Phi$, indeed, induce perturbations in the energy momentum tensor which can be easily expressed in terms of $\delta\phi/\dot{\phi}_0$ and $\Phi$ as
\begin{align}
 \delta T^0_{~0} & = (\rho_{\phi} + P_{\phi})\left[\frac{d}{dt}\left(\frac{\delta\phi}{\dot{\phi}_0}\right) - \Phi \right] - 3H(\rho_{\phi} + P_{\phi})\left(\frac{\delta\phi}{\dot{\phi}_0}\right)\,,\\
\delta T^i_{~i} & = -\left[ \delta T^0_{~0} - 2V_{\rm eff}^{\prime}\dot{\phi}_0\left(\frac{\delta\phi}{\dot{\phi}_0}\right)\right]\,,\\
\delta T^0_{~i} & = (\rho_{\phi} + P_{\phi})\partial_i\left(\frac{\delta\phi}{\dot{\phi}_0}\right)\,,
\end{align}
where there is no sum over $i$ in $\delta T^i_{~i}$, $\delta T^0_{~i} = \delta T_{0i}$ and $\delta T^i_{~j}  = 0$ if $i\neq j$. 
The perturbed (linearized) Einstein field equations become (assuming a flat Universe)
\begin{align}
\frac{\triangle\Phi}{a^2}  -3H\dot{\Phi} - 3H^2\Phi & = \frac{1}{2\bar{M}_{\rm Pl}^2}\delta T^0_{~0}\,,\label{eq:Pert1}\\
\frac{\Phi}{a^2}\left(\dot{a}^2 + 2a\ddot{a}\right) + 4H\dot{\Phi} + \ddot{\Phi} &=  \frac{1}{2\bar{M}_{\rm Pl}^2}\left( \delta T^0_{~0} - 2V_{\rm eff}^{\prime}\delta\phi\right)\,,\label{eq:Pert2}\\
\dot{\Phi} + H\Phi & = \frac{1}{2\bar{M}_{\rm Pl}^2}Z_{\rm eff}\dot{\phi}_0 \delta\phi\,,\label{eq:Pert3}
\end{align}
where the Laplacian on the left-hand side of eq.\,(\ref{eq:Pert1}) is given by $\triangle\Phi = (1/\sqrt{\kappa})\partial_i[\sqrt{\kappa}\kappa^{ij}(\partial_j\Phi)]$, with 
$\kappa$ determinant of the matrix $\kappa_{ij}$ defined at the beginning of section\,\ref{sec:HubbleGauge}.
If we subtract eq.\,(\ref{eq:Pert1}) from eq.\,(\ref{eq:Pert2}) we find (after using $\dot{H} = \ddot{a}/a - H^2$, the EoM\,(\ref{eq:EoMt}) to eliminate 
$V^{\prime}_{\rm eff}$, eq.\,(\ref{eq:Pert3}) solved for $\delta\phi$ and the definition of the Hubble slow-roll parameters $\epsilon_H$ and $\eta_H$) the equation for the gravitational potential 
is
\begin{align}
\ddot{\Phi} + \dot{\Phi}H(1+2\eta_H) + 2\Phi H^2(\eta_H - \epsilon_H) -\frac{\triangle\Phi}{a^2} = 0\,.
\end{align}
This is analogue to the usual equation for the gravitational potential in the standard case (see, e.g., ref.\,\cite{Riotto:2002yw}) with all the non-trivial dependence on $Z_{\rm eff}$ encoded in the Hubble slow-roll parameters $\epsilon_H$, eq.\,(\ref{eq:HubbleEps}), and $\eta_H$, eq.\,(\ref{eq:EtaGaugeInv}). 
To proceed further, we derive an evolution equation for $\mathcal{R}$, after introducing instead of $\Phi$ the new variable $\Xi$ defined by 
$H\Xi = 2\bar{M}_{\rm Pl}^2\Phi a$. 
Eqs.\,(\ref{eq:Pert1}-\ref{eq:Pert3}) can be recast in the form
\begin{align}
\dot{\mathcal{R}} &= \frac{CH}{a}\triangle\Xi\,,~~~~~~~~~{\rm with}~~C \equiv \frac{H}{a^2(\rho_{\phi} + P_{\phi})}\,,\label{eq:R1} \\
\dot{\Xi} & = \frac{a(\rho_{\phi} + P_{\phi})}{H^2}\mathcal{R}\,.\label{eq:R2}
\end{align}
If we take the time-derivative of eq.\,(\ref{eq:R1}) we find, after using  eq.\,(\ref{eq:R2}) and going to the conformal time $\tau$ defined by $dt = a d\tau$
\begin{align}
\frac{d^2\mathcal{R}}{d\tau^2} - \frac{1}{CH}\frac{d(CH)}{d\tau}\frac{d\mathcal{R}}{d\tau} - \triangle\mathcal{R} = 0\,.
\end{align}
We now rescale $\mathcal{R}$ according to 
\begin{align}
\mathcal{R}(\tau,x^i) = -\frac{u(\tau,x^i)}{z(\tau)}\,,
\end{align}
and we choose the conformal time dependence of $z$ in such a way to eliminate the first derivative of $u$
\begin{align}\label{eq:MSyet}
-\frac{1}{z}\frac{d^2 u}{d\tau^2} + \frac{du}{d\tau}\underbrace{\Bigg[
\frac{2}{z^2}\frac{dz}{d\tau} + \frac{1}{z(CH)}\frac{d(CH)}{d\tau}
\Bigg]}_{ \overset{!}{=}\,0} + \frac{u}{z^2}\Bigg[
\frac{d^2 z}{d\tau^2} - \frac{2}{z}\left(\frac{dz}{d\tau}\right)^2 - \frac{1}{(CH)}\frac{d(CH)}{d\tau}\frac{dz}{d\tau}
\Bigg] + \frac{\triangle u}{z} = 0\,.
\end{align}
This condition translates into a differential equation for $z$ of the form $dz/z =-(1/2)d(CH)/CH$ that can be easily solved giving $z = (CH)^{-1/2}$.
All in all, we find from eq.\,(\ref{eq:MSyet}) with the condition $z = (CH)^{-1/2}$ the Mukhanov-Sasaki equation in Fourier space\footnote{For a generic quantity $g(t,\mathbf{x})$ we have
\begin{align}
g(t,\mathbf{x}) = \int \frac{d\mathbf{k}}{(2\pi)^3}  e^{i \mathbf{x}\cdot \mathbf{k}} g_{\mathbf{k}}(t)\,.
\end{align}
}
\begin{align}\label{eq:uk}
\frac{d^2 u_{\mathbf{k}}}{d\tau^2} + \left(
k^2 - \frac{1}{z}\frac{d^2 z}{d\tau^2}
\right)u_{\mathbf{k}} = 0\,,~~~~~~~~{\rm with}\,\,\,z = a\sqrt{Z_{\rm eff}}\,\frac{d\phi_0}{dN_e}\,,
\end{align}
where the difference with respect to the standard result is the presence of the factor $\sqrt{Z_{\rm eff}}$ in the definition of $z$. 
We now apply the slow-roll limit. At the first order in the Hubble slow-roll parameters $\epsilon_H$ and $\eta_H$ we find
\begin{align}
\frac{1}{z}\frac{d^2 z}{d\tau^2} = a^2 H^2 (1 + \epsilon_H -\eta_H)(2-\eta_H) +  a^2 H^2\frac{d}{dN}(\epsilon_H -\eta_H) \simeq a^2 H^2(2+ 2\epsilon_H - 3\eta_H)\,,
\end{align}
that is again equal to the standard result but, crucially, with the Hubble slow-roll parameters given by eq.\,(\ref{eq:HubbleEps}) and eq.\,(\ref{eq:EtaGaugeInv}) when expressed in terms of $V_{\rm eff}$ and $Z_{\rm eff}$.
If we now set 
 \begin{align}\label{eq:MSeq}
\frac{1}{z}\frac{d^2 z}{d\tau^2} = \frac{1}{\tau^2}\left(\nu^2 - \frac{1}{4}\right)\,,
\end{align}
we find for $\nu$ the first-order expression\footnote{We include slow-roll corrections to the expression of the scale factor in terms of the conformal time, namely $a(\tau) = -(1+\epsilon_H)/(H\tau)$.}
 \begin{align}
\nu \simeq \frac{3}{2} + 2\epsilon_H - \eta_H\,.
\end{align}
We can now exploit the known solutions of eq.\,(\ref{eq:MSeq}) in terms of $\nu$\,\,\cite{Riotto:2002yw}. On super-Hubble scales $k\ll aH$ we have
\begin{equation}
\lim_{k\ll aH}\left|\frac{u_{\mathbf{k}}}{z}\right| = \frac{H^2}{\sqrt{Z_{\rm eff}}\dot{\phi}_0\sqrt{2k^{3/2}}}\left(\frac{k}{aH}\right)^{-2\epsilon_H + \eta_H}\,.
\end{equation}
The primordial power spectrum for $\mathcal{R}$ at the linear order in the slow-roll parameters can then be obtained as
  \begin{align}\label{eq:PSgauge}
  \mathcal{P}_{\mathcal{R}}(k) = \frac{k^3}{2\pi^2}\left|\frac{u_{\mathbf{k}}}{z}\right|_{k\ll aH} = 
  \underbrace{\frac{H^2}{8\pi^2 \epsilon_H \bar{M}_{\rm Pl}^2}}_{= A_s}\left(\frac{k}{aH}\right)^{-4\epsilon_H + 2\eta_H} = A_s\left(\frac{k}{aH}\right)^{n_s - 1}\,,
  \end{align}
  with $n_s = 1-4\epsilon_H +2\eta_H$.
  In terms of the potential slow-roll parameters we find the conventional result $n_s = 1+2\eta_V - 6\epsilon_V$ but now with $\epsilon_V$ and $\eta_V$ defined as in eq.\,(\ref{eq:GaugeInvEpsilon}) and eq.\,(\ref{eq:GaugeInvEta}).
  Gauge-fixing invariance of eq.\,(\ref{eq:PSgauge}) is guaranteed by the fact that the expression for $\mathcal{P}_{\mathcal{R}}(k)$  involves gauge-invariant quantities, as discussed in section\,\ref{sec:HubbleGauge} and \ref{sec:SlowRollPar}.
  
 As mentioned in the introduction, there are well-motivated cases in which eq.\,(\ref{eq:PSgauge})
  is not enough to fully describe the power spectrum at small scales in the presence of an ultra slow-roll phase\,\cite{Chongchitnan:2006wx,Motohashi:2017kbs,Germani:2017bcs,Ballesteros:2017fsr}. In such cases, one has to solve eq.\,(\ref{eq:uk}) exactly. 
  It is useful to rewrite the dynamics of $u_{\mathbf{k}}$ as a function of the number of $e$-folds. 
  We find\footnote{To derive this equation, we used 
  \begin{align}
  \frac{d}{dN_e}(\epsilon_H - \eta_H) = -\frac{\bar{M}_{\rm Pl}^2(3-\epsilon_H)V_{\rm eff}^{\prime\prime}}{V_{\rm eff}Z_{\rm eff}} + 
  \frac{(-3+\eta_H)Z_{\rm eff}^{\prime}(d\phi_0/dN_e)}{2Z_{\rm eff}}\,.
   \end{align}
  }
\begin{align}\label{eq:Fulluk}
 \frac{d^2u_{\mathbf{k}}}{dN_e^2} + (1-\epsilon_H)\frac{du_{\mathbf{k}}}{dN_e} + 
 \left[
 \frac{k^2}{a^2 H^2} + (1+\epsilon_H - \eta_H)(\eta_H -2) - \frac{d}{dN_e}(\epsilon_H - \eta_H)
 \right]u_{\mathbf{k}} = 0\,.
  \end{align} 
 This is again equivalent to the standard result but with the Hubble slow-roll parameters given by eq.\,(\ref{eq:HubbleEps}) and eq.\,(\ref{eq:EtaGaugeInv}).
 Eq.\,(\ref{eq:Fulluk}) is manifestly gauge-fixing invariant.
 
\subsection{The tensor-to-scalar ratio}

Tensor fluctuations decouple from scalar ones. We can, therefore, apply the standard computations. The  gauge-invariant tensor amplitude satisfies the equation 
\begin{align}
\frac{d^2v_{\mathbf{k}}}{d\tau^2} + \left(k^2 - \frac{1}{a}\frac{d^2 a}{d\tau^2}\right)v_{\mathbf{k}} = 0\,,
\end{align}
and, as done in eq.\,(\ref{eq:MSeq}), we can again introduce the index $\nu_t$ by computing explicitly the conformal-time derivative  $d^2 a/d\tau^2$ (including again the slow-roll corrections to the expression of $a(\tau)$). 
At the first order in the slow-roll Hubble parameters, we  find the standard result $\nu_t \simeq 3/2 + \epsilon_H$. 
This gives the usual tensor power spectrum on super-Hubble scales 
\begin{align}
\mathcal{P}_t(k) = \underbrace{\frac{8}{\bar{M_{\rm Pl}^2}}\left(\frac{H}{2\pi}\right)^2}_{=\,A_t}\left(\frac{k}{aH}\right)^{-2\epsilon_H} = A_t\left(\frac{k}{aH}\right)^{n_t}\,.
\end{align}
This expression is gauge-fixing invariant with the slow-roll Hubble parameter $\epsilon_H$ given by eq.\,(\ref{eq:HubbleEps}).

\section{Gauge independence of inflationary observables: An explicit example}
\label{sec:CHI}

The goal of this section is to construct the derivative expansion of the effective action in eq.\,(\ref{eq:QuadraticS}) in the one-loop approximation in the context of an explicit model relevant for inflation.

\subsection{The model and the quantum effective action}
\label{sec:Model}
We consider the Abelian Higgs model with a quartic potential and a minimal coupling to gravity.
The action is
\begin{equation}\label{eq:MainActionJord}
\mathcal{S} = \int d^4 x\sqrt{-g}\Bigg[
-\frac{\bar{M}_{\rm Pl}^2}{2}
\mathcal{R}(g) \underbrace{-\frac{1}{4}F_{\mu\nu}F^{\mu\nu} + 
(D_{\mu}\Phi)^*(D^{\mu}\Phi) - \lambda(\Phi^*\Phi)^2}_{\equiv \mathcal{L}_{\rm matter}} + \mathcal{L}_{\rm g.f.} + \mathcal{L}_{\rm ghosts}
\Bigg]\,,
\end{equation}
where $\Phi \equiv (\varphi_1 + i \varphi_2)/\sqrt{2}$ is a complex scalar field,  $F_{\mu\nu}$ is the field strength of the gauge field $A_{\mu}$ and 
the gauge covariant derivative acts on $\Phi$ according to $D_{\mu}\Phi = \partial_{\mu}\Phi + i(g/2)A_{\mu}\Phi$ with $g$ the gauge coupling.
$g$ is the (non-dynamical) background metric. 
The infinitesimal gauge transformations with respect to the space-time dependent infinitesimal gauge parameter $\delta\Theta(x)$ are
\begin{equation}\label{eq:GaugeTr}
\delta A_{\mu}(x) = \partial_{\mu}\delta\Theta(x)\,,~~~~~~
\delta\Phi(x) = -\frac{ig}{2}\delta\Theta(x)\Phi(x) \Longrightarrow
\left\{
\begin{array}{c}
 \delta{\rm Re}[\Phi(x)] =  +\frac{g}{2}\delta\Theta(x){\rm Im}[\Phi(x)]  \\
     \\
 \delta{\rm Im}[\Phi(x)] =  -\frac{g}{2}\delta\Theta(x){\rm Re}[\Phi(x)]
\end{array}
\right.
\end{equation}
Explicitly, we have
\begin{align}\label{eq:Master1}
\mathcal{L}_{\rm matter} = \underbrace{-\frac{1}{4}F_{\mu\nu}F^{\mu\nu}}_{\equiv \mathcal{L}_{\rm gauge}^{\rm kin}} +
\underbrace{\frac{1}{2}
\left[
ig\left(\partial_{\mu}\Phi^*\right)A^{\mu}\Phi + h.c.
\right] + \frac{g^2}{4}A_{\mu}A^{\mu}\Phi^* \Phi}_{\equiv \mathcal{L}_{A\Phi}}
+ \underbrace{\left(\partial_{\mu}\Phi^*\right)\left(\partial^{\mu}\Phi\right) 
 -\lambda\left(\Phi^*\Phi\right)^2}_{\equiv \mathcal{L}_{\Phi}}\,,
\end{align}
while  $\mathcal{L}_{\rm g.f.}$ and $\mathcal{L}_{\rm ghosts}$ are, respectively, 
the gauge fixing and ghost Lagrangian that are added to $\mathcal{L}_{\rm matter}$  to quantize the theory. 
Our goal is the computation of the one-loop effective action, and we shall use the background field method. 
We expand the full quantum field $\varphi_i$ around a (generically non-constant) 
background field configuration $\phi_i(x)$ plus a pure quantum fluctuation $\pi_i(x)$. 
Without loss of generality, we align the background along the real direction
 \begin{equation}\label{eq:BFM}
\varphi_1(x) = \phi(x) + h(x)\,,~~~~~\varphi_2(x) = G(x)\,,
\end{equation}
where $h(x)$ and $G(x)$ are the Higgs and Goldstone quantum fluctuations.
We do not consider the presence of a background field for the $A_{\mu}(x)$ which is a pure quantum fluctuation.
At the lowest order in the expansion proposed in eq.\,(\ref{eq:BFM}) we just have the classical (tree-level) dynamics 
 described by the action 
\begin{align}\label{eq:BGL}
\mathcal{S}_{\rm tree} = \int d^4x\left[
\frac{1}{2}(\partial_{\mu}\phi)(\partial^{\mu}\phi) - V_0(\phi)\right]\,,~~~~~{\rm with}\,\,V_0(\varphi) \equiv \frac{\lambda}{4}(\varphi_1^2 + \varphi_2^2)^2 =
 \frac{\lambda}{4}(\varphi_i\varphi_i)^2\,,
\end{align}
where sum over pairs of identical indices is understood.
At one loop, the effective action can be obtained  by means of the method of steepest descent, and one finds the simple expression (see appendix\,\ref{app:A} for details)
\begin{align}\label{eq:MasterTraceMain}
\mathcal{S}_{\rm 1\,loop} = \sum_{{\rm quanta}}i\eta\,{\rm Tr}\left\{\log\left[\Delta^{-1}_{ab}(x,y)\right]\right\} = 
\sum_{{\rm quanta}}i\eta\int d^4x d^4y\,{\rm tr}\left\{\log\left[\Delta^{-1}_{ab}(x,y)\right]\right\}\delta^{(4)}(x-y)
\,,
\end{align}
where the sum is extended to all quantum fields  appearing in the quantum action, and we have $\eta = 1/2$ for bosons and $\eta = -1$ for fermions and ghost fields $c$, $\bar{c}$.  
In eq.\,(\ref{eq:MasterTrace}), $\Delta^{-1}_{ab}(x,y)$ is the inverse propagator in position space, and it is given by the second functional derivative 
\begin{align}\label{eq:InverseProp}
\Delta^{-1}_{ab}(x,y) = \frac{\delta^2\mathcal{S}}{\delta \mathcal{Q}_a(x)\delta \mathcal{Q}_b(y)}\,,~~~~~~ {\rm with}\,\,\mathcal{Q}_a = \{A_{\mu},h,G, c,\bar{c}\}\,.
\end{align}
In order to extract the inverse propagator, therefore, we need to expand the Lagrangian density  in eq.\,(\ref{eq:MainActionJord}) at the quadratic order in the quantum fluctuations. 
Let us consider first the term $\mathcal{L}_{\Phi}$.
We can extract the inverse propagator for the quantum fluctuations $h$, $G$
\begin{align}\label{eq:InversePropagator}
\Delta_{ij}^{-1}(x,y) = \left[-\delta_{ij}\partial_{x}^2 - U_{ij}(x) \right]\delta^{(4)}(x-y)\,,~~~~~~
U_{ij} \equiv \left.\frac{\partial^2 V(\varphi)}{\partial\varphi_i \partial \varphi_j}\right|_{\{\varphi_1,\varphi_2\} = \{\phi,0\}}
= \left(
\begin{array}{cc}
 3\lambda\phi(x)^2 &  0   \\
 0 &  \lambda\phi(x)^2
\end{array}
\right)\,,
\end{align}
where $\partial_x \equiv \partial/\partial x^{\mu}$ and where the scalar mass matrix $U_{ij}$ depends on $x$ through the space-time dependence of the non-constant background 
$\phi(x)$. 
This is, however, not yet our final result since we need to discuss the rest of the model. We consider first the interaction 
Lagrangian $\mathcal{L}_{A\Phi}$ in eq.\,(\ref{eq:Master1}).
 \begin{align} \label{eq:Rest}
 \mathcal{L}_{A\Phi} = \frac{1}{2}
\left[
ig\left(\partial_{\mu}\Phi^*\right)A^{\mu}\Phi + h.c.
\right] + \frac{g^2}{4}A^2|\Phi|^2 &= 
\frac{g}{2}A^{\mu}\varphi_{i}\varepsilon_{ij}(\partial^{\mu}\varphi_j) + \frac{g^2}{8}
A^{\mu}A_{\mu}(\varphi_i\varphi_i)
\,,
\end{align} 
with $\varepsilon_{ij} = ((0,1),(-1,0))$ the Levi-Civita symbol in two dimensions. Expanding around the background at the quadratic order in the field fluctuations, we find
\begin{align}\label{eq:GaugeMixing}
 \mathcal{L}_{A\Phi}^{(2)} =
 \frac{g}{2}A^{\mu}\left[
 \phi(\partial_{\mu}G) - G(\partial_{\mu}\phi)
 \right] + \frac{g^2\phi^2}{8}A^{\mu}A_{\mu}\,.
 \end{align}
 The first term is a gauge boson-scalar mixing that we shall cancel with the gauge-fixing condition. The third term is a background-dependent mass for the gauge boson
 \begin{align}
 M_A^2 \equiv \frac{g^2\phi^2}{4}\,.
 \end{align}
  We opt for a background $R_{\xi}$ gauge, and we define the gauge-fixing Lagrangian
   \begin{align}
 \mathcal{L}_{\rm g.f.} & = -\frac{1}{2\xi}\mathcal{F}^2 = -\frac{1}{2\xi}\left(
 \partial_{\mu}A^{\mu} - \frac{g\xi}{2}\phi G
 \right)^2\label{eq:GugeFixingF} \\
 & = -\frac{1}{2\xi}(\partial_{\mu}A^{\mu})^2 - \frac{g^2\xi}{8}\phi^2 G^2 
 -\frac{g}{2}A^{\mu}\phi (\partial_{\mu}G) - \frac{g}{2}A^{\mu}\left(\partial_{\mu}\phi
 \right)G\,,\label{eq:GaugeFixingExt}
 \end{align}
 where in the second line an integration by parts is implied.
 The first term in eq.\,(\ref{eq:GaugeFixingExt}) is the usual kinetic correction to the gauge boson propagator, and we find the standard result in $R_{\xi}$ gauge
  \begin{align}\label{eq:InverseGaugeProp}
 \Delta_{\mu\nu}^{-1}(x,y) = \left[
 g_{\mu\nu}\partial_x^2 - \left(1-\frac{1}{\xi}\right)\partial_{\mu}\partial_{\nu} + M_A^2 g_{\mu\nu}
 \right]\delta^{(4)}(x-y)\,.
 \end{align}
  The second term in eq.\,(\ref{eq:GaugeFixingExt}) contributes to the scalar mass matrix. This means that 
 we have an additional  contribution  to the inverse propagator in eq.\,(\ref{eq:InversePropagator}), defined now according to
\begin{align}\label{eq:InversePropagator2}
[\Delta_{\xi}^{-1}(x,y)]_{ij} = \left\{-\delta_{ij}\partial_x^2 - U^{\xi}_{ij}(x)\right]\delta^{(4)}(x-y)\,,~~~~~{\rm with}\,\,
U^{\xi}_{ij} \equiv \left.\frac{\partial^2 V(\varphi)}{\partial\varphi_i \partial \varphi_j}\right|_{\{\varphi_1,\varphi_2\} = \{\phi,0\}} +
\frac{\xi g^2\phi^2}{4}\,.
\end{align}
 The third term in eq.\,(\ref{eq:GaugeFixingExt}) cancels the mixing found in eq.\,(\ref{eq:GaugeMixing}). From what is left in the sum of eq.\,(\ref{eq:GaugeMixing}) and eq.\,(\ref{eq:GaugeFixingExt}), we are left with the mixed interaction 
 \begin{align}\label{eq:MixingInteractions}
 \mathcal{L}_{\rm mix} & =  - gA^{\mu}(\partial_{\mu}\phi)G \,.
 \end{align}
This term is irrelevant for the computation of the effective potential in the limit in which the background is constant (since it vanishes) but it becomes important when one includes the background dynamics.
Finally, we compute the ghost Lagrangian by taking, as customary, the infinitesimal variation of  $\mathcal{L}_{\rm g.f.}$ with respect to a $U(1)$ gauge transformation
\begin{align}
 \mathcal{L}_{\rm ghosts} &= \int d^4y \bar{c}(x)\mathcal{M}(x,y)c(y)\,,\\
 \mathcal{M}(x,y) &= \frac{\delta\mathcal{F}(x)}{\delta\Theta(y)} = \left[\partial_{\mu}\partial^{\mu} + 
 \frac{\xi g^2}{4}\phi(\phi + h)\right]\delta^{(4)}(x-y)\,,
\end{align}
where the function $\mathcal{F}$ is implicitly defined in eq.\,(\ref{eq:GugeFixingF}).
The
gauge transformations are given in eq.\,(\ref{eq:GaugeTr}).
We find the ghost mass 
\begin{align}\label{eq:GhostMass}
M_c^2 = \frac{\xi  g^2 \phi^2}{4}= \xi M_A^2\,.
\end{align}
At the quadratic order in the ghost fields, we extract the inverse ghost propagator
\begin{align}\label{eq:GhostInvProp}
\Delta^{-1}_{c\bar{c}}(x,y) = (\partial_{\mu}\partial^{\mu} + M_c^2)\delta^{(4)}(x-y)\,.
\end{align}
We now have all the ingredients that are needed to compute the one-loop effective action in eq.\,(\ref{eq:MasterTraceMain}).  
This computation is almost trivial if one works in the limit of constant background field (that recovers the usual Coleman-Weinberg effective potential) but it becomes more involved if the space-time dependence of the background is included. 
We refer to appendix\,\ref{app:C} for the explicit computation, and we shall discuss the result in the following sections. 
We consider the one-loop effective action in the form 
\begin{align}
\mathcal{S}_{\rm 1\,loop}  & = \int d^4 x \mathcal{L}_{\rm 1\,loop} =\int d^4 x \bigg\{
\frac{1}{2}\underbrace{\left[
1 + Z_{\rm eff}^{(1)}
\right]}_{= Z_{\rm eff}}(\partial_{\mu}\phi)(\partial^{\mu}\phi) -  \underbrace{\bigg[V_0 + V_{\rm eff}^{(1)}\bigg]}_{= V_{\rm eff}}
\bigg\} \,.
\end{align}
 
 \subsubsection{The effective potential}\label{sec:EffPot}
 
 We find the usual Coleman-Weinberg effective potential (hereafter $\kappa \equiv 1/(4\pi)^2$)
 \begin{align}\label{eq:MasterVeff}
 V_{\rm eff}^{(1)} = \frac{\kappa}{4}&\left[
 3M_A^4\left(\log\frac{M_A^2}{\mu^2} - \frac{5}{6}\right) - M_c^4\left(\log\frac{M_c^2}{\mu^2} - \frac{3}{2}\right) +
 M_h^4\left(\log\frac{M_h^2}{\mu^2} - \frac{3}{2}\right) + M_G^4\left(\log\frac{M_G^2}{\mu^2} - \frac{3}{2}\right)
 \right]\,,
 \end{align}
 with the background-dependent masses summarized in table\,\ref{tab:Masses}.
 
 \begin{table}[htp]
	\begin{center}
	\begin{adjustbox}{max width=1\textwidth}
		\begin{tabular}{||c||c||c||}\hline
		{\color{VioletRed4}{\textbf{Gauge boson}}}	& {\color{VioletRed4}{\textbf{Scalar sector}}} & {\color{VioletRed4}{\textbf{Ghost}}}  \\ \hline\hline
 \multirow{3}{*}{$M_A^2 = \frac{g^2\phi^2}{4}$}     & 
 \multirow{2}{*}{$ M_h^2 = 
 3\lambda\phi^2$} &  \multirow{3}{*}{$M_{c}^2 = \frac{\xi g^2\phi^2}{4} = \xi M_A^2$}		\\
  & &  \\
  & \multirow{2}{*}{$ M_{G}^2 =
 \lambda\phi^2 + \frac{\xi g^2\phi^2}{4} = 
 \lambda\phi^2
 +  \xi M_A^2$} &  \\
  & & \\\hline
		\end{tabular}
		\end{adjustbox}
	\end{center}\vspace{-0.5cm}\caption{
	{\color{VioletRed4}{Mass spectrum of the model discussed in section\,\ref{sec:Model}.}}
	}
	\label{tab:Masses}
\end{table}
 We are now in the position to check the Nielsen identity at the lowest order in the derivative expansion, eq.\,(\ref{eq:ZerothNiel}). 
 To this end, we need to compute the functional $K$ in eq.\,(\ref{eq:Nielsen}) in the limit of constant background field. 
 The explicit expression of $K$ is given by
 \begin{align}\label{eq:ExplK}
 K[\phi(x), \xi] = \frac{i}{2}\int d^4 y\left\langle
 \bar{c}(y)\left[
 \mathcal{F}(y) -2\xi\frac{\partial \mathcal{F}(y)}{\partial\xi}
 \right]\frac{\delta h(x)}{\delta\Theta(x)}c(x)
  \right\rangle_{\rm 1PI}\,,
 \end{align}
 where $\mathcal{F}$ is gauge-fixing functional defined in eq.\,(\ref{eq:GugeFixingF}), and
 where the subscript 1PI indicates that we only have to include one-particle irreducible graphs.  
 Notice that in eq.\,(\ref{eq:ExplK}) it enters the variation $\delta h(x)$ under an infinitesimal gauge transformation $\delta\Theta(x)$ of the field 
 $h(x)$ with background $\phi(x)$.
 The corresponding gauge transformation is given in eq.\,(\ref{eq:GaugeTr}), and we have  $\delta h(x)/\delta\Theta(x) = (g/2)G(x)$. 
 The integrand function in eq.\,(\ref{eq:ExplK}), therefore, reads
  \begin{align}
 \bar{c}(y)\left[
 \mathcal{F}(y) -2\xi\frac{\partial \mathcal{F}(y)}{\partial\xi}
 \right]\frac{\delta h(x)}{\delta\Theta(x)}c(x) =  \frac{g}{2}\bar{c}(y)
 \left[
 \partial_{\mu}A^{\mu}(y) + \frac{g\xi}{2}\phi(y)G(y)
 \right]G(x)c(x)\,.
 \end{align} 
 We compute the expectation value in eq.\,(\ref{eq:ExplK}) diagrammatically.  
 First of all, notice that the Nielsen identity in eq.\,(\ref{eq:ZerothNiel}) at one loop takes the form 
 \begin{align}\label{eq:NIlowest}
 \xi\frac{\partial V^{(1)}_{\rm eff}}{\partial\xi} &= C^{(1)}V_{0}^{\prime}\,,
\end{align}
 where on the right-hand side we have the tree-level potential since the functional $K$ (and its constant background limit $C$) starts at one loop. 
 In the constant background limit, we only have one contribution
\begin{align}
C^{(1)}
 =
\resizebox{25mm}{!}{
\parbox{30mm}{
\begin{tikzpicture}[]
\node (label) at (0,0)[draw=white]{ 
       {\fd{2.75cm}{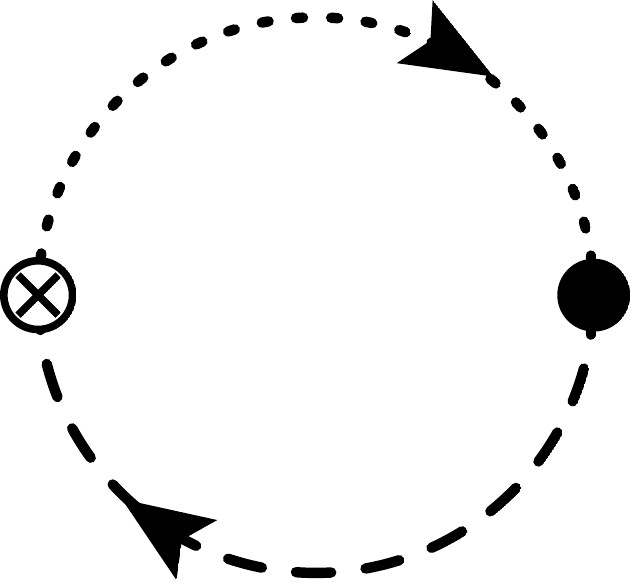}} 
      };
\node[anchor=east] at (-1.2,0.28) {$x$};
\node[anchor=west] at (1.25,0.28) {$y$};
\node[anchor=north] at (0.,1.1) {$c$};
\node[anchor=south] at (0,-1.1) {$G$};
\end{tikzpicture}
}}~~
 & =  -\frac{i\xi g^2\phi}{8}\int d^4y D_{c}(y-x)D_{G}(x-y) = \frac{ig^2\xi\phi}{8}\int \frac{d^4k}{(2\pi)^4}
\frac{1}{(k^2 - M_{c}^2)(k^2 - M^2_{G})}\nn\\ 
&= \frac{\kappa \xi g^2\phi}{8\underbrace{(M^2_{G}-M_{c}^2)}_{= \lambda\phi^2}}\left[
M_{G}^2\left(
\log\frac{M_{G}^2}{\mu^2} - 1
\right)-
M_{c}^2\left(
\log\frac{M_{c}^2}{\mu^2} - 1
\right)
\right]\,,\label{eq:FinalC}
\end{align}
where we introduced the position-space propagator
\begin{equation}
D_{w}(x-y)=
\resizebox{25mm}{!}{
\parbox{30mm}{
\begin{tikzpicture}[]
\node (label) at (0,0)[draw=white]{ 
       {\fd{2.cm}{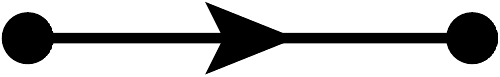}} 
      };
\node[anchor=east] at (-0.8,0.3) {$y$};
\node[anchor=east] at (0.4,0.34) {$k$};
\node[anchor=west] at (.8,0.3) {$x$};
\end{tikzpicture}
}}
=
 \int\frac{d^4 k}{(2\pi)^4}\frac{ie^{-ik\cdot(x-y)}}{(k^2 - M_{w}^2 +i\epsilon)}\,,~~~~~{\rm with}~~\int d^4 x e^{ik\cdot  x} = (2\pi)^4\delta^{(4)}(k)\,.
\end{equation}
On the left-hand side of eq.\,(\ref{eq:NIlowest}) we have
\begin{align}
\xi\frac{\partial V^{(1)}_{\rm eff}}{\partial\xi}  = \frac{\kappa\xi}{2}\left[
M_G^2\frac{\partial M_G^2}{\partial\xi}\left(
\log\frac{M_{G}^2}{\mu^2} - 1
\right)  - M_c^2 \frac{\partial M_c^2}{\partial\xi}\left(
\log\frac{M_{c}^2}{\mu^2} - 1
\right)
\right]\,,
\end{align}
and eq.\,(\ref{eq:NIlowest}) is trivially verified with $\partial M_G^2/\partial\xi = \partial M_c^2/\partial\xi = g^2\phi^2/4$. 
Before proceeding, let us consider the parametric scaling $\lambda\sim O(\kappa g^4)$.  
This is a useful assumptions for both practical and conceptual reasons (even though it makes less transparent the usual fixed-order loop expansion\,\cite{Andreassen:2014eha}).
In this limit we have $\kappa M_A^4 \sim O(\kappa g^4)$, $M^4_{h} \sim \lambda^2\sim O(\kappa^2 g^8) \to 0$ and $M^2_{G}\to M^2_{c} = \xi M_A^2$. The effective potential reduces to 
\begin{align}\label{eq:SimplifiedPot}
V_{{\rm eff},\lambda\sim O(\kappa g^4)} = \frac{\kappa g^4\phi^4}{4}\left[\frac{\lambda}{\kappa g^4} + \frac{3}{16}\left(
\log\frac{g^2\phi^2}{4\mu^2} - \frac{5}{6}
\right)\right]\,,~~~~~~~~\frac{\partial}{\partial\xi}\left[V_{{\rm eff},\lambda \sim O(\kappa g^4)}\right] = 0\,,
\end{align}
and we do not have any explicit gauge dependence.  
This is a useful limit because in eq.\,(\ref{eq:SimplifiedPot}) the radiative correction is as large as the tree-level piece. 
This means that  in this limit radiative corrections have the chance to modify the minimum of the tree-level potential (at $\phi=0$). 
The effective potential, indeed, develops a minimum at $\phi = v_{\phi}$ if
\begin{align}\label{eq:DimTra}
\left.\frac{\partial V_{{\rm eff},\lambda\sim O(\kappa g^4)}}{\partial\phi}\right|_{\phi = v_{\phi}} = 0~~~~~~\Longrightarrow~~~~~~
\lambda =\frac{\kappa g^4}{16}\left(
1-3\log\frac{g^2v_{\phi}^2}{4\mu^2}
\right)\,.
\end{align}
At the scale $\mu = v_{\phi}$, we have 
\begin{align}\label{eq:DimTra1}
\lambda = \frac{\kappa g^4}{16}\left(
1- 3\log\frac{g^2}{4}
\right)\,.
\end{align}
This equation must be regarded as a condition on the renormalization group flow of the dimensionless coupling constants $g$ and $\lambda$: The minimum occurs at the scale $\mu = v_{\phi}$ 
where the condition 
in eq.\,(\ref{eq:DimTra1}) is met. This is one (well-known) paradigmatic example of dimensional transmutation\,\cite{Coleman:1973jx}. The original theory is scale-invariant because there is no dimensionful parameter, and 
the scale $\mu$ was introduced as the renormalization scale.
The theory, however, develops a mass scale $v_{\phi}$ at which a specific relationship among couplings -- in our case eq.\,(\ref{eq:DimTra1}) -- occurs.
The term dimensional transmutation refers to the fact that, instead of using the two dimensionless couplings, the theory can be described 
in terms of  a single coupling and a mass scale $v_{\phi}$, together with a relation that determines the second coupling at the specific scale $v_{\phi}$. 
The Nielsen identity in eq.\,(\ref{eq:NIlowest}) is trivially verified since the tree-level potential is of order $\lambda\sim O(\kappa g^4)$ while we have $C^{(1)} \sim O(\kappa g^2)$ 
(and, consequently,  the right-hand side of eq.\,(\ref{eq:NIlowest}) is of two-loop order $O(\kappa^2 g^6)$).
The same limit in eq.\,(\ref{eq:FinalC}) indeed  gives
\begin{align}
C^{(1)}_{\lambda\sim O(\kappa g^4)} = \frac{\kappa \xi g^2\phi}{8}\log\frac{\xi M_A^2}{\mu^2}\,,
\end{align}
meaning that we expect in this limit an explicit gauge dependence from the corrections to the kinetic term of order $O(\kappa g^2)$ that we shall compute in the next section.
For later use, we also compute the field derivative of $C^{(1)}_{\lambda\sim O(\kappa g^4)}$. We find
\begin{equation}\label{eq:LaterUseC}
[C^{(1)}_{\lambda\sim O(\kappa g^4)}]^{\prime} = \frac{\kappa \xi g^2}{8}\left(2+\log\frac{\xi M_A^2}{\mu^2}\right)\,.
\end{equation}

  \subsubsection{Corrections to the kinetic term}\label{sec:KinTerms}
  
  We find the following sum of terms
 \begin{align}\label{eq:TotalZ}
 Z^{(1)}_{\rm eff} = 
  \underbrace{\left. Z^{(1)}_{\rm eff}\right|_{\rm gauge}}_{{\rm eq}.\,(\ref{eq:ZGauge})} + 
 \underbrace{\left. Z^{(1)}_{\rm eff}\right|_{\rm Higgs}}_{{\rm eq}.\,(\ref{eq:ZHiggs})} + 
 \underbrace{\left. Z^{(1)}_{\rm eff}\right|_{\rm Goldstone}}_{{\rm eq}.\,(\ref{eq:ZGoldstone})} + 
  \underbrace{\left. Z^{(1)}_{\rm eff}\right|_{\rm ghosts}}_{{\rm eq}.\,(\ref{eq:ZGhost})} +  
 \underbrace{\left. Z^{(1)}_{\rm eff}\right|_{\rm mix}}_{{\rm eq}.\,(\ref{eq:Zmix})}\,,
 \end{align} 
 where the corresponding expressions are collected in appendix\,\ref{app:C}.
 We can try to check the Nielsen identity in eq.\,(\ref{eq:FirstNiel}). At one loop, it reads
\begin{align}\label{eq:OneLoopFirstNiels}
\xi\frac{\partial Z^{(1)}_{\rm eff}}{\partial\xi} &=  2[C^{(1)}]^{\prime} - 2D^{(1)}V_{0}^{\prime} + 2[\tilde{D}^{(1)}V_{0}^{\prime}]^{\prime}\,,
\end{align}
where we used again the fact that the functional $K$ (and, therefore, all the coefficients of its derivative expansion) starts at one loop. 
The explicit check of eq.\,(\ref{eq:OneLoopFirstNiels}) requires the knowledge of the coefficients $D^{(1)}$ and $\tilde{D}^{(1)}$.
However, we do not have the corresponding expressions at our disposal (as discussed in section~\ref{sec:Guage},
they are not important for the gauge-invariant formulation of inflation proposed in this paper if one stops at the quadratic order in the derivatives).
Nevertheless, it is still possible to check in a non-trivial way the validity of our computation against eq.\,(\ref{eq:OneLoopFirstNiels})  if we take the limit $\lambda\sim O(\kappa g^4)$. 
In this limit, the tree-level potential is of order $\lambda\sim O(\kappa g^4)$ and the terms $D^{(1)}V_{0}^{\prime}$ and $[\tilde{D}^{(1)}V_{0}^{\prime}]^{\prime}$ in eq.\,(\ref{eq:OneLoopFirstNiels})
sub-leading. We are left with 
\begin{align}\label{eq:OneLoopFirstNielsSimpl}
\xi\frac{\partial Z^{(1)}_{{\rm eff},\lambda\sim O(\kappa g^4)}}{\partial\xi} &=  2[C_{\lambda\sim O(\kappa g^4)}^{(1)}]^{\prime} = 
\frac{\kappa \xi g^2}{4}\left(2 - \log\frac{1}{\xi}+\log\frac{g^2\phi^2}{4\mu^2}\right)\,.
\end{align}
We checked the validity of this identity by computing explicitly the left-hand side by means of eq.\,(\ref{eq:TotalZ}) with $Z^{(1)}_{{\rm eff}}$ that, in the limit $\lambda\sim O(\kappa g^4)$, reads
\begin{align}
Z_{{\rm eff},\lambda\sim O(\kappa g^4)} = 1+ \frac{\kappa g^2}{4}\left[
\xi - \xi\log\frac{1}{\xi} + (3+\xi)\log\frac{g^2\phi^2}{4\mu^2}
\right]\,.
\end{align}
We stress that  verifying this identity is a non-trivial check since on the left-hand side it requires a conspiracy among the $\xi$-dependent ghost, gauge, Goldstone and mixing contributions in eq.\,(\ref{eq:TotalZ}).

\subsection{Few simple applications}\label{sec:Apply}

In this section we shall briefly discuss some of the relations obtained in section\,\ref{sec:Guage} from a more operative point of view.  

\subsubsection{Mathematical consistency}\label{sec:Math}

We check here the mathematical consistency of some of the equations derived in section\,\ref{sec:Guage}. 
\begin{figure}[!ht!]
\begin{center}
$$\includegraphics[width=.42\textwidth]{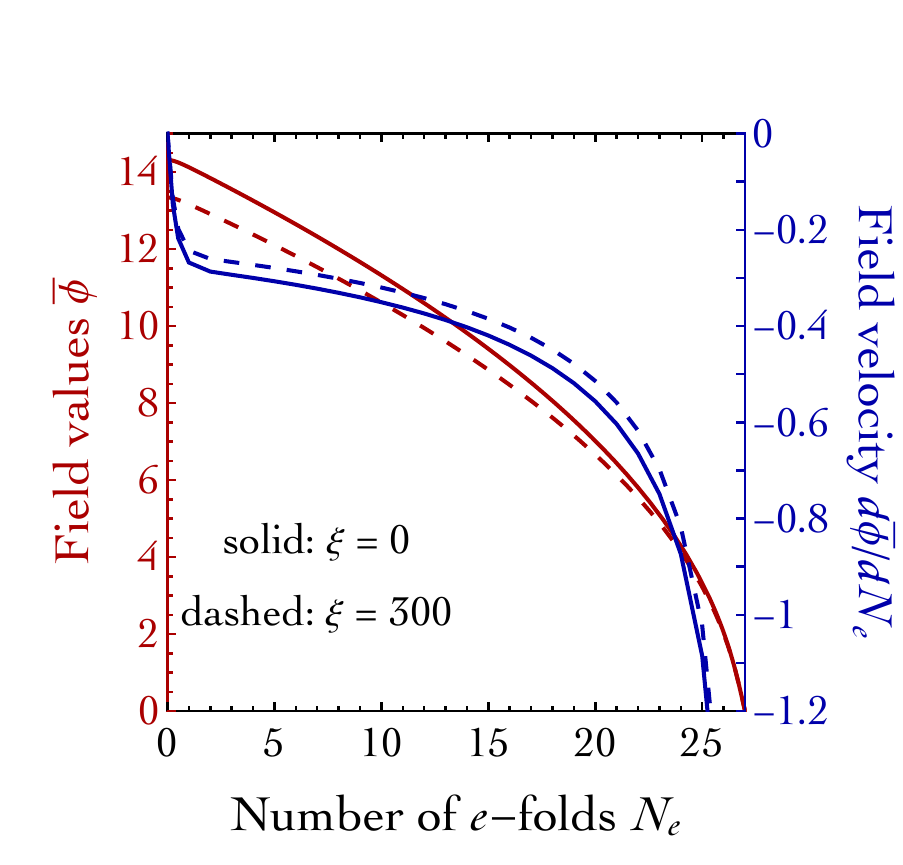}
\qquad\includegraphics[width=.42\textwidth]{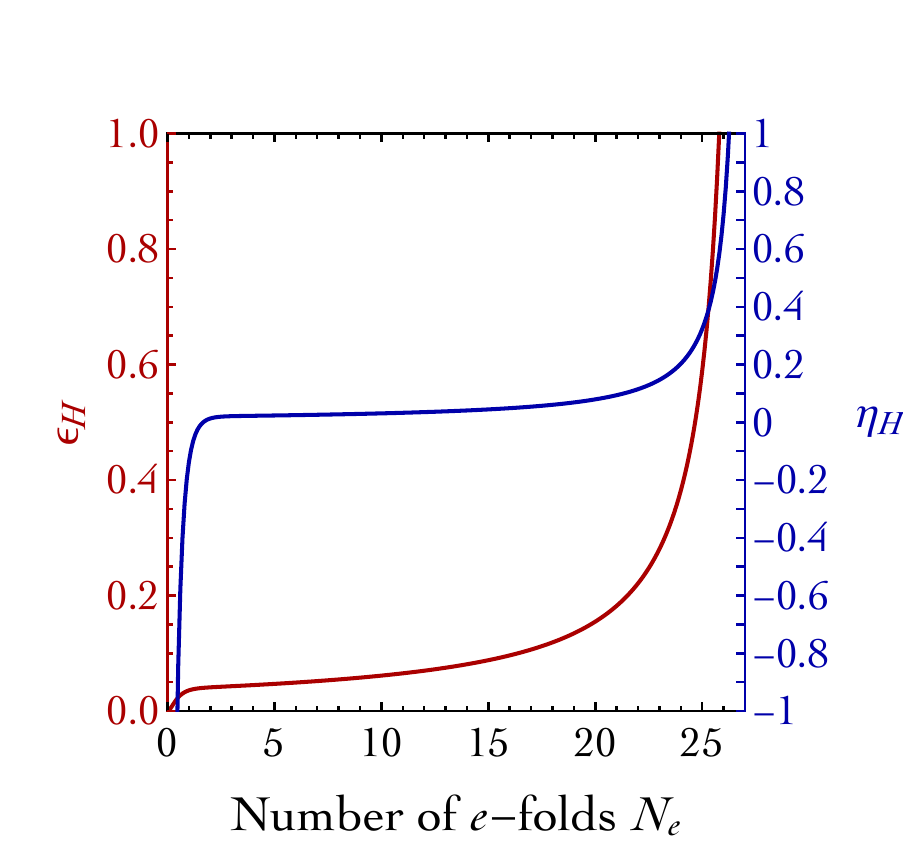}$$
\caption{\em \label{fig:Ka} 
{\color{VioletRed4}{Left panel. Gauge-fixing dependence of field values (red, left-side $y$-axis) and velocities (blue, right-side $y$-axis) as a function of the number of $e$-folds as time variable. 
We show two representative values for the gauge-fixing parameter $\xi$, namely $\xi = 0$ (solid) and $\xi = 300$ (dashed). 
The gauge-fixing dependence in the field values and velocities -- obtained by solving the EoM\,(\ref{eq:EoMN}) for the two analyzed values of $\xi$ -- agree with that dictated by eq.\,(\ref{eq:CheckXiDep}). Right panel. Gauge-fixing independence (solid and dashed lines coincide) of the observable Hubble parameters $\epsilon_H$ (eq.\,(\ref{eq:HubbleEps}), left-side $y$-axis in red) and $\eta_H$ (eq.\,(\ref{eq:EtaGaugeInv}), right-side $y$-axis in blue) as a function of the number of $e$-folds as time variable.
}}
 }
\end{center}
\end{figure}
As representative example, we start with a simple check of the gauge-fixing independence of the Hubble parameter $\epsilon_H$ in its slow-roll limit, eq.\,(\ref{eq:GaugeInvEpsilon}). We write
\begin{align}\label{eq:CheckSlowRollEps}
\xi\frac{d\epsilon_V}{d\xi} = \epsilon_V\left[
-\frac{1}{Z_{\rm eff}}\left(
\xi\frac{\partial Z_{\rm eff}}{\partial\xi} + Z_{\rm eff}^{\prime}\xi\frac{d\phi}{d\xi}
\right) -  \frac{2}{V_{\rm eff}}\left(
\xi\frac{\partial V_{\rm eff}}{\partial\xi} + V_{\rm eff}^{\prime}\xi\frac{d\phi}{d\xi}
\right) +\frac{2}{V^{\prime}_{\rm eff}}\left(
\xi\frac{\partial V^{\prime}_{\rm eff}}{\partial\xi} + V^{\prime\prime}_{\rm eff}\xi\frac{d\phi}{d\xi}
\right) 
\right]\,,
\end{align}
and we work at order $O(\kappa)$ in the loop expansion. We already checked (see eq.\,(\ref{eq:NIlowest}) and the related discussion) that 
$\xi\partial V_{\rm eff}/\partial\xi + V_{\rm eff}^{\prime}\xi d\phi/d\xi = O(\kappa^2)$. 
The last term in eq.\,(\ref{eq:CheckSlowRollEps}), on the contrary, gives
\begin{align}
\frac{2}{V^{\prime}_{\rm eff}}\left(
\xi\frac{\partial V^{\prime}_{\rm eff}}{\partial\xi} + V^{\prime\prime}_{\rm eff}\xi\frac{d\phi}{d\xi}
\right)  =  \frac{\kappa \xi g^2}{4\lambda}\left[
\lambda - \frac{\xi g^2}{4}\log\frac{M_c^2}{\mu^2} + \left(
\lambda + \frac{\xi g^2}{4}
\right)\log\frac{M_G^2}{\mu^2}
\right]
+ O(\kappa^2)\,,
\end{align}
where we used, neglecting higher derivatives terms, $\xi d\phi/d\xi = - C$. Up to this point, the computation is valid for any field configuration $\phi$. If we now consider a solution of the EoM $\bar{\phi}$, 
we can neglect in eq.\,(\ref{eq:FirstNiel}) terms with $V^{\prime}_{\rm eff}$ since they contribute at higher order in the derivative expansion (in the slow-roll limit, $V^{\prime}_{\rm eff}$ counts as $\dot{\phi}$). We find
\begin{align}
-\frac{1}{Z_{\rm eff}}\left(
\xi\frac{\partial Z_{\rm eff}}{\partial\xi} + Z_{\rm eff}^{\prime}\xi\frac{d\phi}{d\xi}
\right)  =  -\frac{\kappa \xi g^2}{4\lambda}\left[
\lambda - \frac{\xi g^2}{4}\log\frac{M_c^2}{\mu^2} + \left(
\lambda + \frac{\xi g^2}{4}
\right)\log\frac{M_G^2}{\mu^2}
\right]
+ O(\kappa^2)\,,
\end{align}
thus implying $\left. \xi d\epsilon_V/d\xi\right|_{\phi = \bar{\phi}} = 0 + O(\kappa^2)$.
The most important point in the discussion of gauge invariance is that field values are not observables quantities, and their $\xi$-dependence satisfies a differential equation of the form given in eq.\,(\ref{eq:Key}). 
As stated below this equation, this property is a consequence of the gauge invariance of the effective action, $\xi d\mathcal{S}_{\rm eff}/d\xi = 0$.
Crucially, when computing observable quantities, the gauge-fixing dependence of the field configuration cancels the explicit $\xi$-dependence of the effective action, and leads to the definition of gauge-independent observables. 
Let us, therefore, check for consistency the validity of eq.\,(\ref{eq:Key}). 
We consider the effective action discussed in section\,\ref{sec:Model}, and we focus on solutions of the EoM, denoted as 
$\bar{\phi}$ in section\,\ref{sec:Guage}. Using the number of $e$-folds as time variable, our goal is to check the validity of the relations
\begin{align}\label{eq:CheckXiDep}
\xi\frac{d}{d\xi}\bar{\phi}(N_{e},\xi) = -C^{(1)}(\bar{\phi},\xi)\,,~~~~~~~~~~~~\xi\frac{d}{d\xi}\frac{d\bar{\phi}(N_{e},\xi)}{dN_e} = -[C^{(1)}(\bar{\phi},\xi)]^{\prime}\frac{d\bar{\phi}(N_{e},\xi)}{dN_e}\,.
\end{align} 
In the left panel of fig.\,\ref{fig:Ka} we show the field profile $\bar{\phi}$ (red, referred to the left-side of the y-axis) and its derivative $d\bar{\phi}/dN_e$ (blue, referred to the right-side of the y-axis) as a function of the number of $e$-folds for different values of the gauge-fixing parameter $\xi$. 
To obtain these solutions, we solved -- for each value of $\xi$ -- the corresponding  EoM\,(\ref{eq:EoMN}). 
We checked that the resulting $\xi$-dependence of  $\bar{\phi}$ and $d\bar{\phi}/dN_e$ (solid versus dashed lines in fig.\,\ref{fig:Ka}) precisely matches the one obtained by solving the differential equations  in eq.\,(\ref{eq:CheckXiDep}). 
Notice that the number of $e$-folds covered by the two solutions does not depend on $\xi$, as discussed in section\,\ref{sec:Ne}, when eq.\,(\ref{eq:CheckXiDep}) is satisfied.
In the right panel of fig.\,\ref{fig:Ka}, we computed the Hubble parameters $\epsilon_H$ and $\eta_H$. 
If properly defined as discussed in section\,\ref{sec:SlowRollPar}, they do not depend on the gauge-fixing parameter as indeed verified in this plot (solid and dashed lines coincide) where the $\xi$-dependence of the field profile $\bar{\phi}$
 is compensated by the explicit $\xi$-dependence of $Z_{\rm eff}$ and $V_{\rm eff}$.
 
\subsubsection{Coleman-Weinberg ``hilltop'' inflation}\label{sec:CWhilltop}

Consider the quantum effective action for the model discussed in the previous section. Thanks to radiative corrections and dimensional transmutation, 
it is possible to have a working inflationary model of the so-called ``hilltop'' type (see ref.\,\cite{Kallosh:2019jnl} for a recent reappraisal of this class of inflationary models in light of the most recent Planck data). 
As we shall see, this also provides  the simplest setup in which discussing gauge-fixing invariance. 
The Coleman-Weinberg mechanism\,\cite{Coleman:1973jx} represents an elegant dynamical realization of the inflaton potential, as first implemented in the seminal paper\,\cite{Linde:1981mu}. 
Consider the one-loop effective potential in eq.\,(\ref{eq:MasterVeff}) by taking the leading-log (LL) approximation
\begin{align}\label{eq:VeffLL}
V_{\rm eff,LL}^{(1)} = \frac{\lambda\phi^4}{4} + \frac{\kappa\phi^4}{4}\left(
\frac{3}{8}g^4 + 20\lambda^2 + \xi \lambda g^2
\right)\log\frac{\phi}{\mu} + \Lambda^4\,.
\end{align}
 The last term is a constant potential added by hand in order to solve the cosmological constant problem and avoid problems related to eternal inflation.
 The value of $\Lambda$ is fixed by requiring that the value of the potential at the minimum where inflation ends is zero.
The LL approximation has the advantage of being independent from the renormalization scheme used.
If we adopt the same approximation, we find the one-loop correction to the kinetic term
\begin{align}\label{eq:ZeffLL}
Z_{\rm eff,LL}^{(1)} = 1 + \frac{\kappa g^2}{2}(3+\xi)\log\frac{\phi}{\mu}\,,
\end{align}
while the one-loop functional coefficient entering in the Nielsen identity takes the form
\begin{align}
C^{(1)}_{\rm LL} = \frac{\kappa\xi g^2}{4}\phi \log\frac{\phi}{\mu}\,.
\end{align}
The part of the effective potential in eq.\,(\ref{eq:VeffLL}) 
 that does not depend on the renormalization scheme
matches  the one used in ref.\,\cite{Barenboim:2013wra} (after one turns off the contribution of the Yukawa coupling of the right-handed neutrinos 
and identifies the gauge coupling $g_X$ with our $g/4$\,\cite{Chun:2013soa}) in which the Landau gauge, $\xi = 0$, was chosen.  
At this level of approximation, implementing the Nielsen identity is rather trivial. 
Consider the field redefinition 
\begin{align}
\frac{d\tilde{\phi}}{d\phi} = \left[Z_{\rm eff,LL}^{(1)}\right]^{1/2}~~~~~~\Longrightarrow~~~~~~\tilde{\phi} \overset{{\rm LL}}{=}\phi\left[
1 + \frac{\kappa g^2}{4}(3+\xi)\log\frac{\phi}{\mu}
\right]\,,
\end{align}
which renders the kinetic term for $\tilde{\phi}$ canonical. In terms of the field $\tilde{\phi}$ the effective potential reads
\begin{align}\label{eq:GaugeInvEffPot}
V_{\rm eff,LL}^{(1)} = \frac{\lambda\tilde{\phi}^4}{4} + \frac{\kappa\tilde{\phi}^4}{4}\left(
\frac{3}{8}g^4 + 20\lambda^2 -3\lambda g^2
\right)\log\frac{\tilde{\phi}}{\mu} + \Lambda^4\,,
\end{align}
with no $\xi$-dependence left. In terms of $\tilde{\phi}$, therefore, the kinetic term is canonical and the effective potential does not depend explicitly on the gauge-fixing parameter.
Let us make clear the connection with the Nielsen identity.
Since we have $\partial\mathcal{S}_{\rm eff}/\partial\xi = 0$, a direct  consequence of eq.\,(\ref{eq:Key}) is that we must have $d\tilde{\phi}/d\xi = 0$. 
It is indeed easy to check that 
\begin{align}\label{eq:notra}
\xi \frac{d\tilde{\phi}}{d\xi} = \xi\frac{\partial\tilde{\phi}}{\partial\xi} + 
\left(\frac{\partial\tilde{\phi}}{\partial\phi}\right)\underbrace{\xi\frac{d\phi}{d\xi}}_{-C^{(1)}_{\rm LL}}  = 
\left(\frac{\kappa \xi g^2}{4}\phi\log\frac{\phi}{\mu}\right) + \left(-\frac{\kappa \xi g^2}{4}\phi\log\frac{\phi}{\mu}\right) + 
O(\kappa^2) = O(\kappa^2)\,,
\end{align}
where we used the $\xi$-dependence of the field $\phi$ dictated by eq.\,(\ref{eq:Key}).
Eq.\,(\ref{eq:notra}) is consistent with our discussion on the Nielsen identity. 
The effective potential in eq.\,(\ref{eq:GaugeInvEffPot}) differs from the one in eq.\,(\ref{eq:VeffLL}) computed in Landau gauge (as in ref.\,\cite{Barenboim:2013wra}).
Before proceeding, let us quickly compare our result with ref.\,\cite{Mooij:2011fi} where the 
quantum effective action for the same model was computed by means of the closed-time-path formalism (in this approach, one 
first computes one loop corrections to the EoM with the tadpole method and then 
obtains the effective action by performing an integration). The effective action obtained in ref.\,\cite{Mooij:2011fi} agrees with our result in eq.\,(\ref{eq:GaugeInvEffPot}). In addition, in this paper we gave an interpretation of the gauge-fixing independence based on the Nielsen identity.

\begin{figure}[!h!]
\begin{center}
	\includegraphics[width=.53\textwidth]{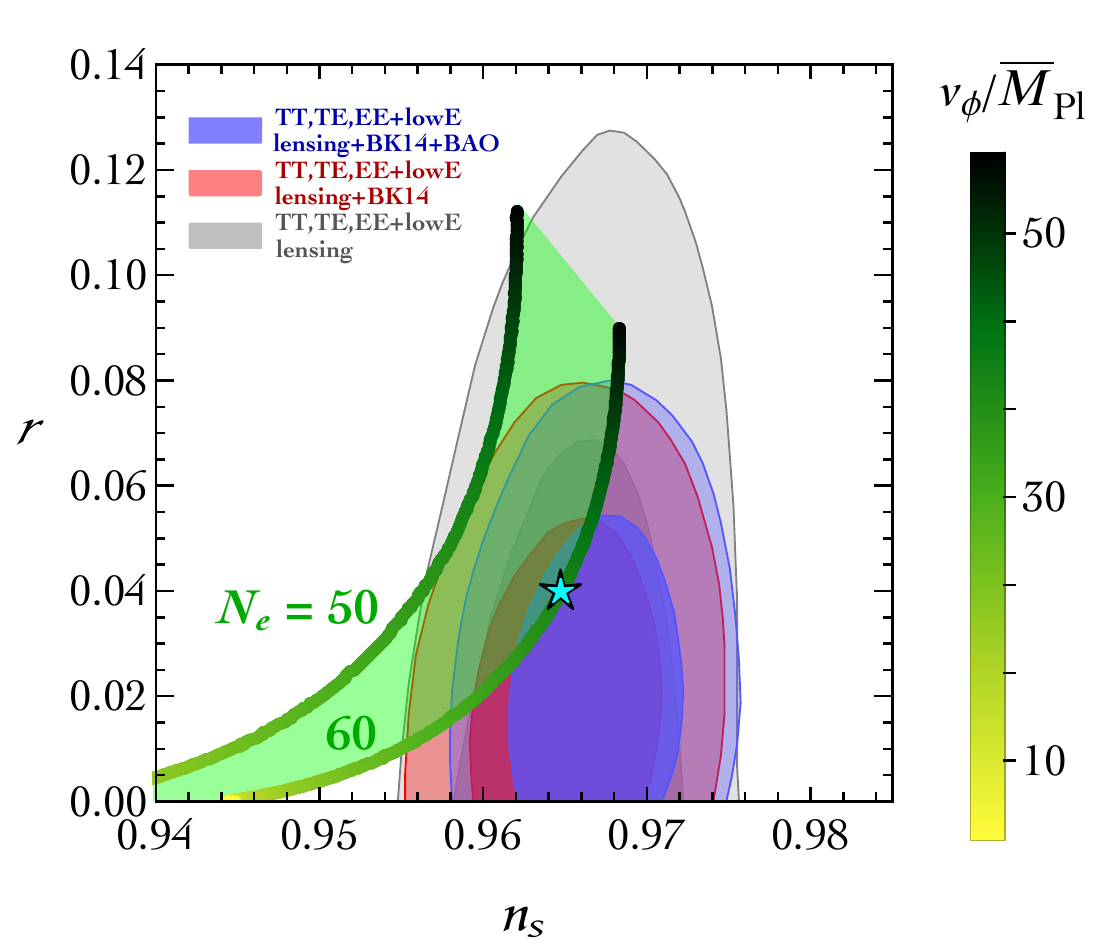}
	\caption{
{\color{VioletRed4}{
Marginalized joint 68\% and 95\% confidence regions  for the spectral index $n_s$ and tensor-to-scalar ratio $r$ in the base $\Lambda$CDM mode using Planck\,TT,TE,EE+lowE (grey), Planck\,TT,TE,EE+lowE+lensing  (red),  and Planck\,TT,TE,EE+lowE+lensing+BAO  (blue). We superimpose the prediction of the gauge-invariant Coleman-Weinberg ``hilltop'' model 
discussed in section\,\ref{sec:CWhilltop}. The bottom (top) boundary of the green region corresponds to $N_e = 60$ ($N_e = 50$) $e$-folds of accelerated expansion. 
Along the two boundaries, the avocado colors mark, according to the corresponding legend bar, the values of the ratio $v_{\phi}/\bar{M}_{\rm Pl}$. 
The cyan star indicates the inflationary solution (with $v_{\phi}/\bar{M}_{\rm Pl} \simeq 23$) whose 
trajectory along the inflaton potential is shown in 
boldface cyan in
eq.\,(\ref{eq:EffPotPlot}). 
}}	
	 \label{fig:CWInfl}}
\end{center}
\end{figure}

For completeness, let us quickly review the inflationary properties of this model. 
Since we assumed that the logarithm is large, the potential must be RG-improved. We have 
(re-labeling $\tilde{\phi}\to \phi$ for simplicity)
\begin{align}\label{eq:RGPote}
V_{\rm eff,LL}^{(1)} = \frac{\lambda(t)}{4}\phi^4 + \Lambda^4\,,
\end{align}
where, as customary, we take for the renormalization scale $\mu = \phi$ (and, consequently, we have $dt = d\log\phi$);
$\lambda(t)$ solves the one-loop RGEs
\begin{align}
\frac{d\lambda(t)}{dt} & = \kappa\left[
\frac{3}{8}g(t)^4 + 20\lambda(t)^2 - 3\lambda(t)g(t)^2
\right]\,,\label{eq:RGE1}\\
\frac{dg(t)}{dt} & = \frac{\kappa}{12}g(t)^3\,.
\end{align}
If the quartic coupling is negative for small $\phi$ and turns positive for large $\phi$,
then the potential in eq.(\ref{eq:RGPote}) develops a minimum close to the point where $\lambda(t)$ crosses zero.
More precisely, using eq.\,(\ref{eq:RGE1}), we see that the minimum of the potential in eq.\,(\ref{eq:RGPote}) occurs at $\phi = v_{\phi}$ where the condition
\begin{align}
\lambda(v_{\phi}) = -\frac{\kappa}{4}\left[
\frac{3}{8}g(v_{\phi})^4 + 20\lambda(v_{\phi})^2 - 3\lambda(v_{\phi})g(v_{\phi})^2
\right]\,,
\end{align}
is verified. Let us indicate with $t_0$ the value of $t$ where $\lambda$ vanishes, $\lambda(t_0) = 0$. 
If we expand around $t_0$ and use eq.\,(\ref{eq:RGE1}), we find
\begin{align}
\lambda(t) \approx (t-t_0)\left.\frac{d\lambda(t)}{dt}\right|_{t=t_0}  =  \frac{3\kappa g_0^4}{8}\log\frac{\phi}{\phi_0}~~~~
\Longrightarrow~~~~V_{\rm eff,LL}^{(1)} \approx \frac{3\kappa g_0^4}{32}\phi^4\log\frac{\phi}{\phi_0} + \Lambda^4\,,
\end{align}
where $g_0\equiv g(t_0)$. 
The minimum of the potential occurs at $v_{\phi} = \phi_0 e^{-1/4}$, and in terms of $v_{\phi}$ we can write
\begin{align}
V_{\rm eff,LL}^{(1)} \approx \frac{3\kappa g_0^4}{32}\phi^4\left[\log\frac{\phi}{v_{\phi}} - \frac{1}{4}\right] + \Lambda^4\,,
\end{align}
which has the parametric structure
\begin{align}\label{eq:EffPotPlot}
V_{\rm eff}(\phi)/V_0
 =
\resizebox{40mm}{!}{
\parbox{21mm}{
\begin{tikzpicture}[]
\node (label) at (0,0)[draw=white]{ 
       {\fd{2.75cm}{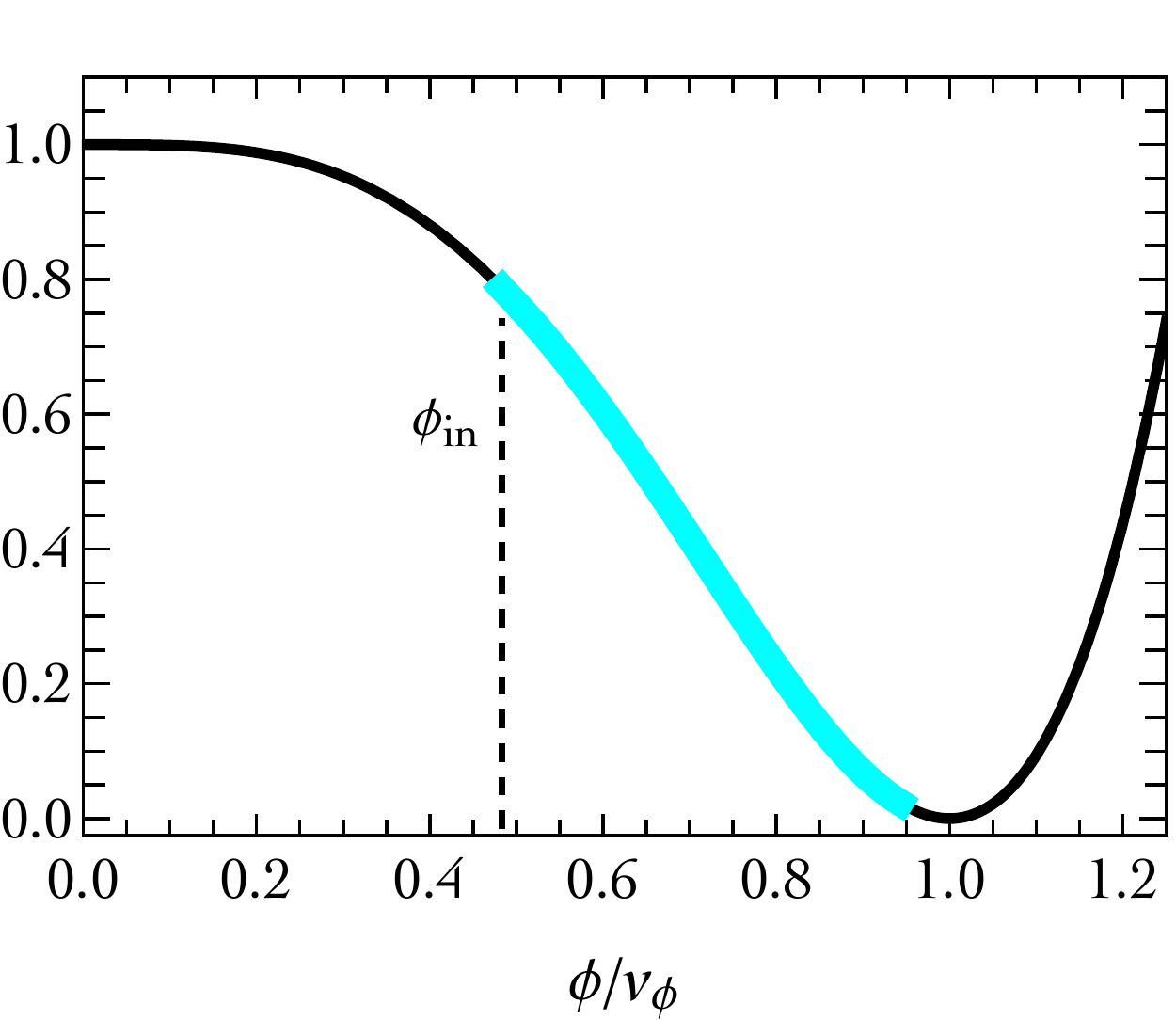}} 
      };
\end{tikzpicture}
}}
 &~~~~~~~~~~~~~ = \left[
 1 + \frac{\phi^4}{v_{\phi}^4}\left(
 2\log\frac{\phi^2}{v_{\phi}^2} - 1
 \right)
 \right]\,,
\end{align}
where the cosmological constant is fixed by the condition $V_{\rm eff}(v_{\phi})= 0$. 
In boldface cyan we indicate the part of the potential where inflation takes place (corresponding to the solution marked with a cyan star in fig.\,\ref{fig:CWInfl}).
The inflationary predictions, as far as the spectral index and the tensor-to-scalar ratio are concerned, 
are shown in fig.\,\ref{fig:CWInfl} (see caption for details).
In order to have enough $e$-folds of accelerated expansion without violating the Planck constraints, 
one needs trans-planckian values  for the symmetry breaking scale, namely $v_{\phi} \gtrsim 30 \bar{M}_{\rm Pl}$. 
However, for the Hubble rate (evaluated at the beginning of inflation by means of the slow-roll approximation) we find
\begin{align}\label{Hubble}
\frac{H_{\rm in}}{\bar{M}_{\rm Pl}} = \left[\frac{V_{\rm eff}(\phi_{\rm in})}{3\bar{M}_{\rm Pl}^2}\right]^{1/2}
 =
\resizebox{40mm}{!}{
\parbox{21mm}{
\begin{tikzpicture}[]
\node (label) at (0,0)[draw=white]{ 
       {\fd{2.75cm}{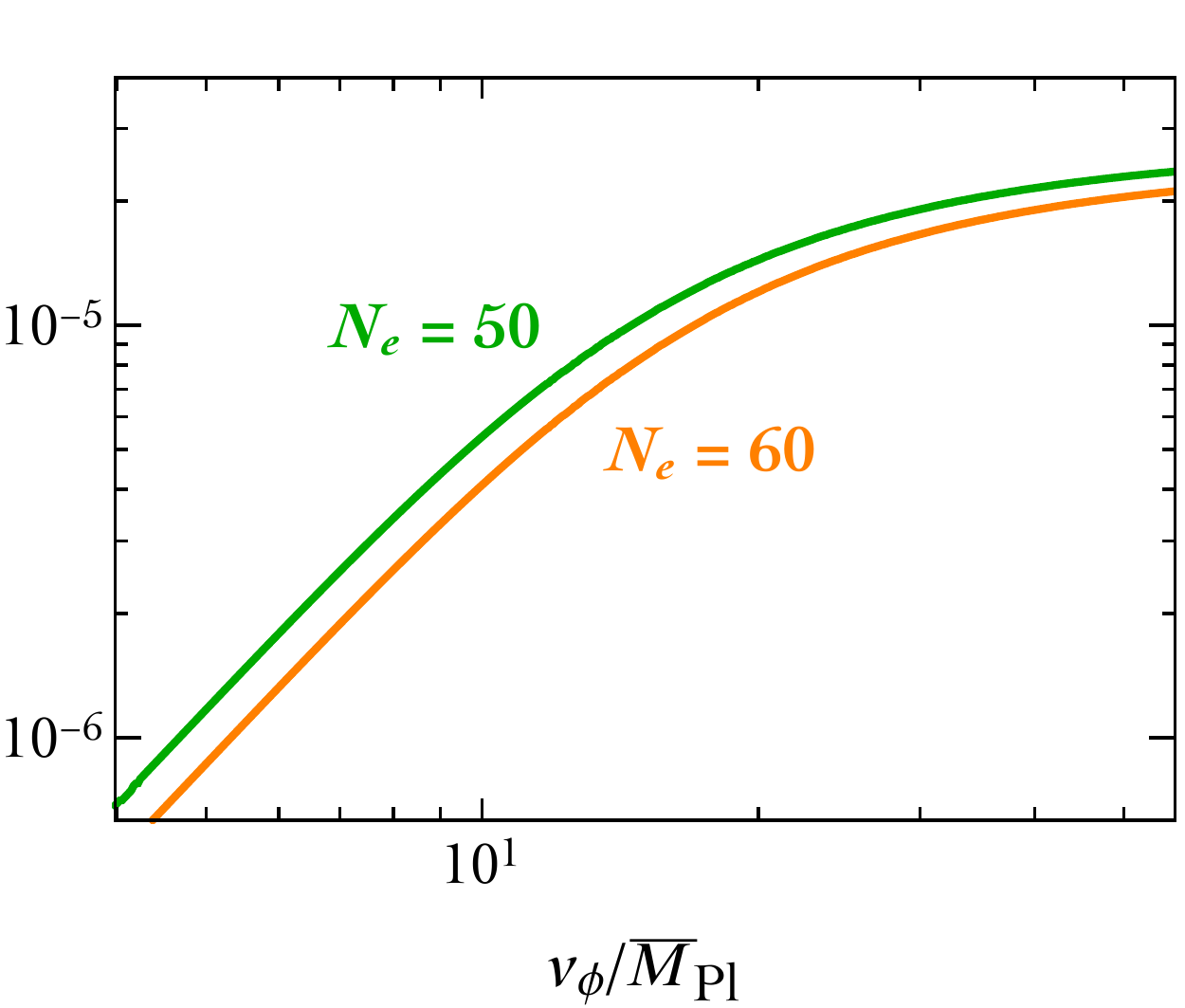}} 
      };
\end{tikzpicture}
}}\,~~~~~~~~~~~~ \simeq 10^{-5}\times \left(\frac{A_s}{10^{-9}}\right)^{1/2}\left(\frac{r}{0.05}\right)^{1/2}\,,
\nn
\end{align}
so that the energy density during inflation remains comfortably  in the sub-planckian regime.

\section{Summing up and conclusions}
\label{sec:Concl}
In this work we elaborated a framework in which cosmological observables related to the inflationary dynamics -- assuming inflation to be described in the context of a gauge theory --  are gauge-fixing independent. 
Our construction, discussed in detail in section\,\ref{sec:Guage}, exploits the power of the Nielsen identity in eq.\,(\ref{eq:Nielsen}) and extends to the case of inflation previous analysis (most notably ref.\,\cite{Espinosa:2016nld}) focused on the gauge-fixing invariance of observable quantities related to the instability of the electroweak scale. 
In section\,\ref{sec:CHI}, we discussed an explicit example based on the abelian Higgs model with a quartic tree-level potential.

In this section, we conclude with a brief discussion of possible future directions to explore. The example studied in section\,\ref{sec:CWhilltop} is extremely simple. 
In the LL approximation, it is indeed straightforward to recognize that the gauge-fixing dependence in eq.\,(\ref{eq:VeffLL}) and in eq.\,(\ref{eq:ZeffLL}) only enters by means 
of the anomalous dimension of the scalar field $\phi$.\footnote{It means that we can write
\begin{align}
V_{\rm eff,LL}^{(1)}  & =  \frac{\lambda\phi^4}{4} + \frac{\phi^4}{4}\left[
\beta_{\lambda}^{(1)} - 4\lambda \gamma_{\phi}^{(1)}
\right]\log\frac{\phi}{\mu}\,,\\
Z_{\rm eff,LL}^{(1)}  & = 1 - 2\gamma_{\phi}^{(1)}\log\frac{\phi}{\mu}\,,
\end{align}
where the one-loop anomalous dimension of $\phi$ is $\gamma_{\phi}^{(1)} = -\kappa g^2(3+\xi)/4$ while $\beta_{\lambda}^{(1)} = \kappa(3g^4/8 + 20\lambda^2 -3\lambda g^2)$ is the one-loop $\beta$ function of the quartic coupling.
} In this respect, the discussion of section\,\ref{sec:CWhilltop} can be considered as a simple situation in which the gauge-fixing dependence is eliminated by a field redefinition that reabsorbs the effect of the anomalous dimension. Beyond the LL approximation, implementing a field redefinition of the same kind -- that is a field redefinition such that the explicit gauge-fixing dependence of the effective potential is eliminated -- is much more complicated\,\cite{Nielsen:2014spa}. 
It is for this reason that in most of the computations performed beyond the LL approximation gauge-fixing independence is implemented just by performing 
the field redefinition that reabsorbs the anomalous dimension. This approximation leads to an effective potential in which the gauge-fixing dependence is only reduced but not eliminated (see the discussion in ref.\,\cite{Espinosa:2015qea}). 
Even though for practical purposes this approximation seems to be enough (see ref.\,\cite{Masina:2018ejw} for a discussion in the context of critical Higgs inflation), insisting on the necessity of performing a field redefinition such to eliminate the gauge-fixing dependence of the effective potential seems, at least at the conceptual level, not the best way to go.
On the contrary, the approach developed in section\,\ref{sec:Guage} (along the line of ref.\,\cite{Espinosa:2016nld} but  in the context of inflation) in which the gauge-fixing dependence of the effective potential is kept intact but observables are defined in a gauge-independent way, 
could reveal to be much more helpful.
It would be, therefore, interesting to apply the formalism developed in  section\,\ref{sec:Guage} to more complicated, but still phenomenologically relevant, cases. 
Let us try to motivate further a possible computation in this direction. Consider the case in which the effective potential features, because of radiative corrections, the presence of a stationary inflection point.
\begin{figure}[!htb!]
\begin{center}
	\includegraphics[width=.35\textwidth]{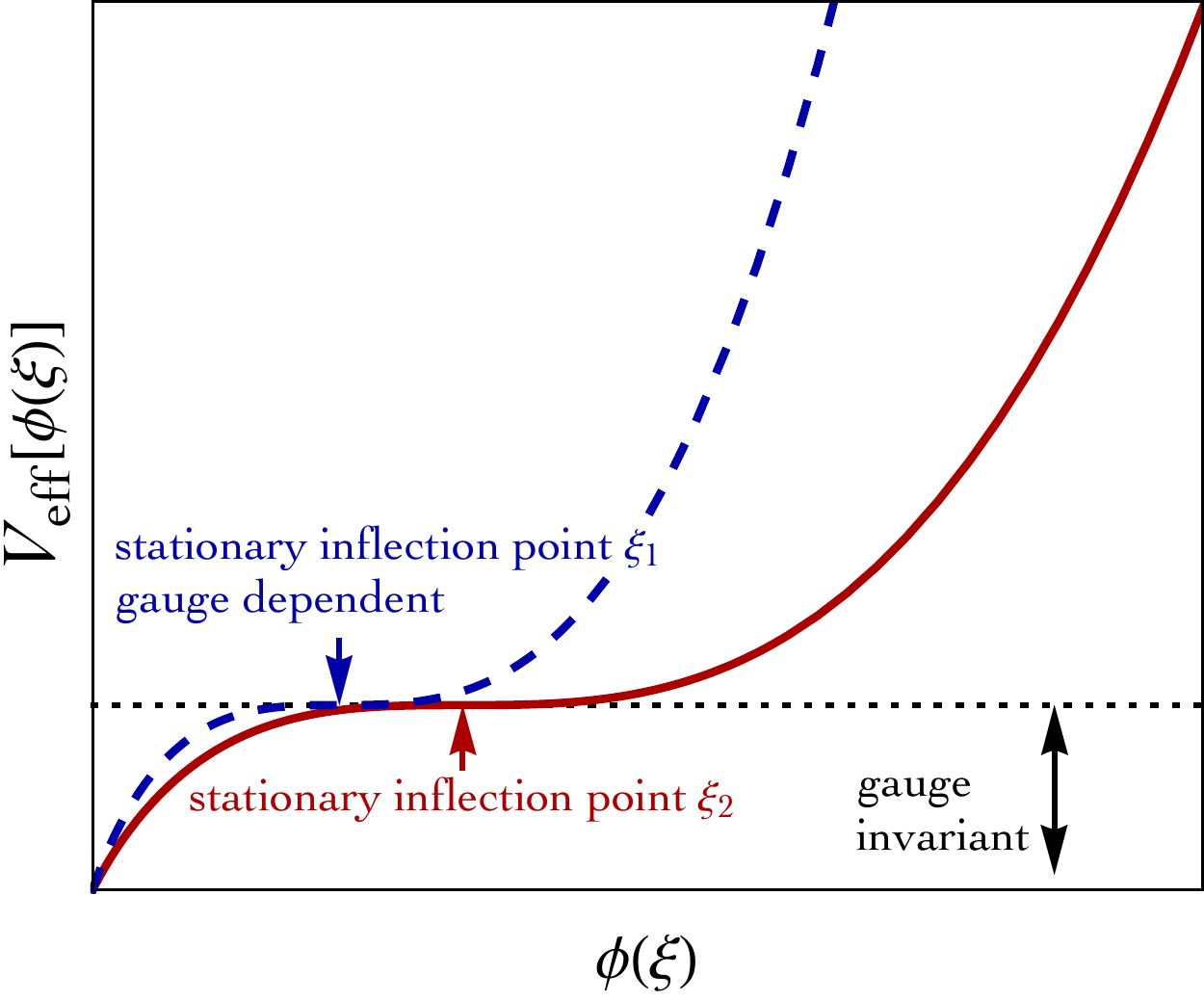}
	\caption{
	{\color{VioletRed4}{
	Qualitative illustration of the 
gauge-fixing dependence of the effective potential in the presence of a stationary inflection point (defined by the two conditions $dV_{\rm eff}/d\phi = 0$ and $d^2V_{\rm eff}/d\phi^2 = 0$). 
The value of the effective potential at the extrema does not depend on the gauge-fixing parameter. The position in field space of the stationary inflection point, on the contrary, does depend on $\xi$.
}}
	 \label{fig:InflectionPointPlot}}
\end{center}
\end{figure}
Qualitatively, we expect the situation illustrated in fig.\,\ref{fig:InflectionPointPlot}. 
The value of the effective potential at the stationary inflection point (horizontal dotted line) is a gauge-invariant quantity (see eq.\,(\ref{eq:ZerothNiel})). 
However, the position in field space of the stationary inflection point does depend on the gauge-fixing parameter $\xi$ (vertical single arrows). 
Inflaton potentials that are characterized by the presence of a stationary inflection point generated by radiative corrections could leave observable imprints in the power spectrum of comoving curvature perturbations at small scales, and  
a prominent example is the possibility to create a population of primordial black holes (that could account for the totality of dark matter in our Universe in the mass window $10^{17} \lesssim M_{\rm PBH}\,[\,{\rm g}\,] \lesssim 10^{21}$). 
However, spurious changes introduced by some residual gauge-dependence -- like the one described qualitatively in fig.\,\ref{fig:InflectionPointPlot}  -- could inficiate the observable predictions of these models (especially if one bears in mind that observables like the abundance of primordial black holes depends exponentially from the  power spectrum of curvature perturbations). 
Conceptually, this situation is quite relevant for the implementation of our formalism in particular if one needs to go beyond the LL approximation and consider, 
for instance, the two-loop RG improvement of the one-loop Coleman-Weinberg potential (as discussed, for instance, in ref.\,\cite{Ballesteros:2015noa} in the context of inflationary models featuring a radiative pleteau).
The biggest obstacle that prevents the immediate applicability of the formalism  developed in section\,\ref{sec:Guage}, however, 
is that, in order to have a working inflationary model that is not ruled out by CMB observations, one needs to flatten the potential in fig.\,\ref{fig:InflectionPointPlot} at large field values. 
This goal can be easily  achieved by introducing a sizable non-minimal coupling to gravity. 
The drawback of this procedure is that the resulting theory in the Einstein frame is no longer renormalizable, and care must be taken when computing radiative corrections (see the discussion in ref.\,\cite{George:2013iia}). 
Furthermore, the scalar field space in the Einstein frame turns out to be a Riemannian manifold, and  a covariant implementation of the background field method  is needed for the correct computation of the corrections to the kinetic term. 
Despite these technical difficulties, given the possible phenomenological relevance of this class of models for the early Universe cosmology, the explicit implementation of a gauge-fixing invariant formalism might be worth a try.
\vspace{0.2cm}
\begin{figure}[!h!]
\begin{center}
	\includegraphics[width=.005\textwidth]{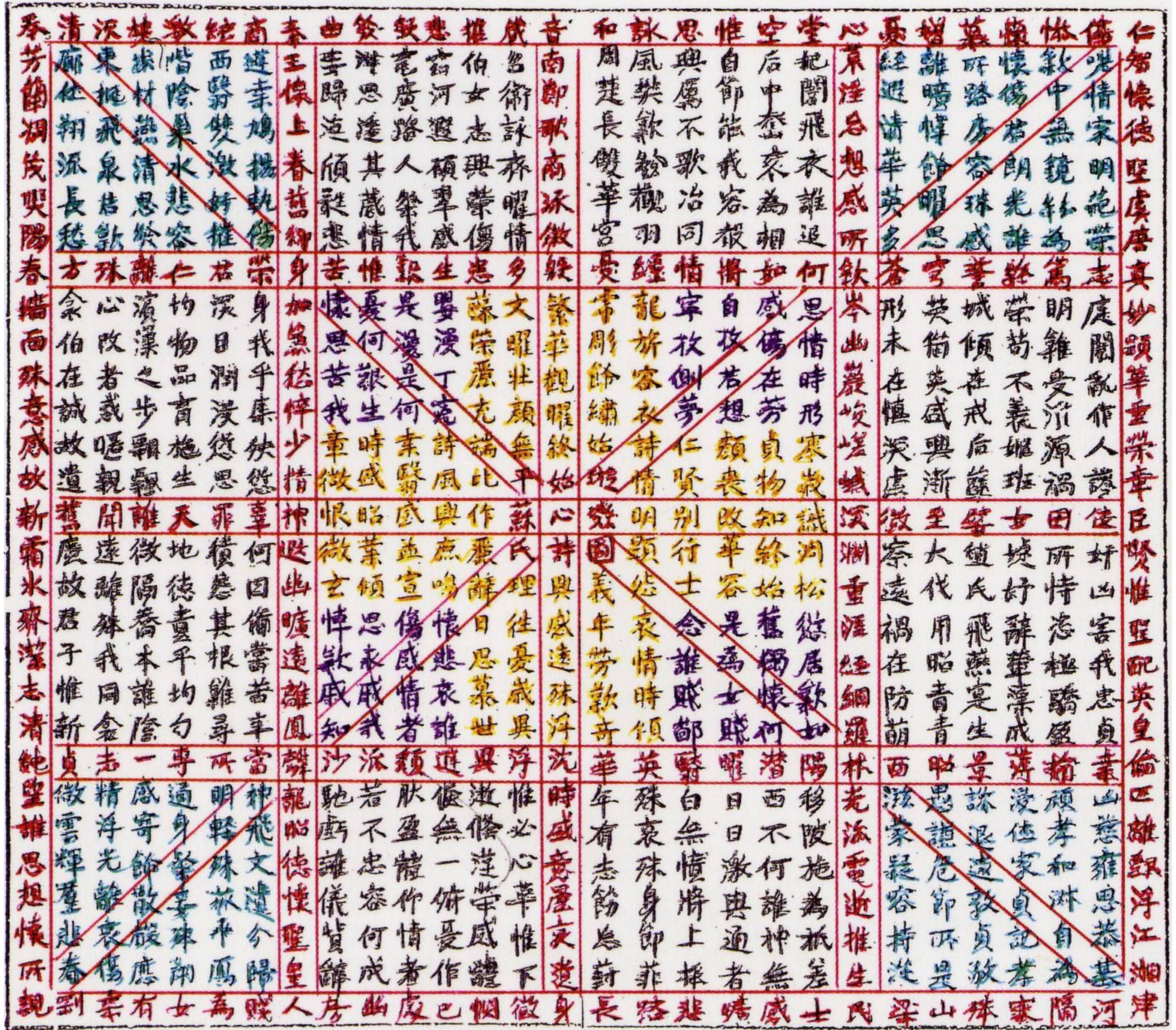}
\end{center}
\end{figure}
\vspace{-0.75cm}
\acknowledgments

I thank G.~Ballesteros, M.~Serone, P.~Serpico and M.~Taoso  for discussions.
This work is partially supported by the MIUR under contract 2017\,FMJFMW (``{\it {\color{VioletRed4}{New avenues in strong dynamics}}},'' PRIN\,2017) and by the INFN grant ``{{\color{VioletRed4}{{ SESAMO}}} -- 
{{\color{VioletRed4}{{S}}}inergi{{\color{VioletRed4}{{ E}}} di {{\color{VioletRed4}{{ SA}}}pore e {{\color{VioletRed4}{{ M}}}ateria {{\color{VioletRed4}{{ O}}}scura.''

\appendix
\section{Paralipomena}
\label{app:A}

Most of the material in this appendix is devoted to the explicit computation of the one-loop effective action for the abelian Higgs model, and it complements the discussion in section\,\ref{sec:CHI}.
In this computation, we shall use the background R$_{\xi}$ gauge.

%

\subsection{Explicit computation of the one-loop effective action}
\label{app:C}

At one loop, the effective action can be obtained  by means of the method of steepest descent, and one finds the simple expression
\begin{align}\label{eq:MasterTrace}
\mathcal{S}_{\rm 1\,loop} = \sum_{{\rm quanta}}i\eta\,{\rm Tr}\left\{\log\left[\Delta^{-1}_{ab}(x,y)\right]\right\} = 
\sum_{{\rm quanta}}i\eta\int d^4x d^4y\,{\rm tr}\left\{\log\left[\Delta^{-1}_{ab}(x,y)\right]\right\}\delta^{(4)}(x-y)
\,,
\end{align}
where the sum is extended to all quantum fields appearing in the quantum action, and we have $\eta = 1/2$ for bosons and $\eta = -1$ for fermions and ghosts. 
In eq.\,(\ref{eq:MasterTrace}),  we isolated from the functional trace Tr the space-time integration (the 
symbol tr denotes the remaining trace over internal degrees of freedom, if any).
In the limit of constant background field, the computation of eq.\,(\ref{eq:MasterTrace}) is trivial and gives the usual Coleman-Weinberg effective potential. 
In the presence of a non-constant background field, the computation is more complicated. 
It is still possible to apply elegant functional methods, as explained in detail in refs.\,\cite{Fraser:1984zb,Aitchison:1984tn,Aitchison:1985pp,Chan:1985ny,Cheyette:1985ue,Gaillard:1985uh,Binetruy:1988nx}.
However, in order to compute the corrections to the kinetic term we instead follow an intuitive diagrammatic approach\,\cite{Fraser:1984zb}.  
We have
\begin{align}\label{eq:DefZ}
Z_{\rm eff}^{(1)} = \sum_{{\rm q}} \left.Z_{\rm eff}^{(1)}\right|_{\rm q} =\sum_{{\rm q}} \left.\frac{d\Sigma_{hh}^{(1)}}{dp^2}\right|^{\rm q}_{p^2 = 0}\,,
\end{align}
 where $\Sigma_{hh}^{(1)}$ is the one-loop self-energy of the quantum field $h$ and the sum over ``q'' takes into account all possible quantum fields running in the loop.\footnote{As already discussed in ref.\,\cite{Garny:2012cg}, 
 in the $R_{\xi}$ gauge derivatives with respect to the background field $\phi$ are not related to diagrams with external $h$ fields.
 To account for this issue, it is possible to replace in the gauge fixing term 
 the background field $\phi$ with $\phi + \tilde{h}$ with $\tilde{h}$ treated as an external field. 
 Consequently, eq.\,(\ref{eq:DefZ}) is replaced by 
 \begin{align}\label{eq:DefZfull}
Z_{\rm eff}^{(1)} = \sum_{{\rm q}} \left.Z_{\rm eff}^{(1)}\right|_{\rm q} =\sum_{{\rm q}}
\left(
\left.\frac{d\Sigma_{hh}^{(1)}}{dp^2}\right|^{\rm q}_{p^2 = 0}+
\left.\frac{d\Sigma_{h\tilde{h}}^{(1)}}{dp^2}\right|^{\rm q}_{p^2 = 0}+
\left.\frac{d\Sigma_{\tilde{h}h}^{(1)}}{dp^2}\right|^{\rm q}_{p^2 = 0}+
\left.\frac{d\Sigma_{\tilde{h}\tilde{h}}^{(1)}}{dp^2}\right|^{\rm q}_{p^2 = 0}
\right)
\,.
\end{align}
We implement this method in our computation even though, for the sake of simplicity, we omit the notation used 
in eq.\,(\ref{eq:DefZfull}).
 }
 Consider first the case in which we have the gauge boson $A$ running in the loop.
 We use dimensional regularization in $d=4-2\epsilon$ dimensions. We find
 \begin{align}
 \left.i\Sigma_{hh}^{(1)}\right|^{\rm gauge}
 &=
\resizebox{25mm}{!}{
\parbox{30mm}{
\begin{tikzpicture}[]
\node (label) at (0,0)[draw=white]{ 
       {\fd{2.75cm}{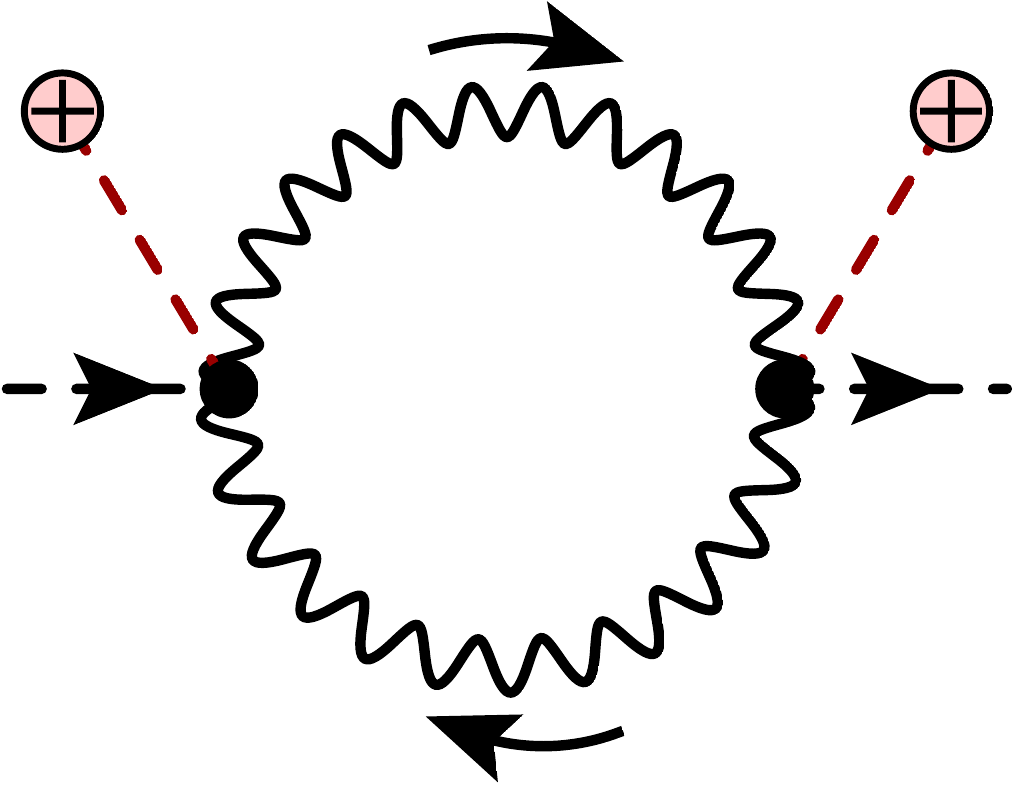}} 
      };
\node[anchor=east] at (-1.,0.25) {$p$};
\node[anchor=north] at (0.5,1.5) {$k+p$};
\node[anchor=south] at (-0.5,-1.2) {$k$};
\node[anchor=east] at (-1.,-0.4) {$h$};
\node[anchor=east] at (-1.,1.2) {{\color{rossos}{$\phi$}}};
\node[anchor=east] at (1.55,1.2) {{\color{rossos}{$\phi$}}};
\node[anchor=west] at (.9,-0.4) {$h$};
\end{tikzpicture}
}}
= -\frac{g^2M_A^2}{2}\int\frac{d^4 k}{(2\pi)^4}g^{\mu\rho}g^{\nu\sigma}P_{\mu\nu}^{(\xi)}(k+p,M_A)P^{(\xi)}_{\rho\sigma}(k,M_A)\nn\\
&= \frac{ig^2}{8(4\pi)^2M_A^2}\left\{2(1-\xi)M_A^2\left[A_0(M_A^2) - A_0(\xi M_A^2)\right] 
+ \left(12M_A^4 -4p^2M_A^2 + p^4 \right)B_0(p^2,M_A^2,M_A^2)
\right.\nn\\
&\left.\hspace{2.5cm}
+\left[
-2M_A^4(1-\xi)^2 + 4(1+\xi)M_A^2p^2 -2p^4
\right]B_0(p^2,M_A^2,\xi M_A^2)\right.\nn\\
&\left.\hspace{2.5cm}
+\left[
4M_A^4\xi^2 -4\xi M_A^2p^2 + p^4
\right]B_0(p^2,\xi M_A^2,\xi M_A^2)
\right\}\,,\label{eq:LoopA}
\end{align}
where  the propagator of the gauge boson is (see eq.\,(\ref{eq:InverseGaugeProp}))
\begin{equation}
P_{\mu\nu}^{(\xi)}(k,M_A) = \frac{-i}{k^2 - M_A^2}\left[
g_{\mu\nu} - (1-\xi) \frac{k_{\mu}k_{\nu}}{k^2 - \xi M_A^2}
\right]\,.
\end{equation}
The scalar one- and two-point integrals $A_0$ and $B_0$ in eq.\,(\ref{eq:LoopA}) are defined in appendix\,\ref{app:B}. 
There are no UV divergences, and we find the kinetic correction
\begin{equation}\label{eq:ZGauge}
\left.Z_{\rm eff}^{(1)}\right|_{\rm gauge} =\left.\frac{d\Sigma_{hh}^{(1)}}{dp^2}\right|^{\rm gauge}_{p^2 = 0} = \frac{\kappa g^2}{24}
\left[
15 + 11\xi - \frac{18\xi}{(1-\xi)}\log\frac{1}{\xi}
\right]\,.
\end{equation}
The limit $\xi = 0$ reproduces the known result in Landau gauge, that is $\left.Z_{\rm eff}^{(1)}\right|^{\xi = 0}_{\rm gauge} = 5\kappa g^2/8$ (see ref.\,\cite{Espinosa:2016uaw}). 
The Feynman gauge limit $\xi = 1$ gives $\left.Z_{\rm eff}^{(1)}\right|^{\xi = 1}_{\rm gauge} = \kappa g^2/3$. 
We now consider the case in which we have the Higgs running in the loop. The corresponding one loop two-point diagram is (including a symmetric factor $1/2$)
  \begin{align}
 \left.i\Sigma_{hh}^{(1)}\right|^{\rm Higgs}
 &=
\resizebox{25mm}{!}{
\parbox{30mm}{
\begin{tikzpicture}[]
\node (label) at (0,0)[draw=white]{ 
       {\fd{2.75cm}{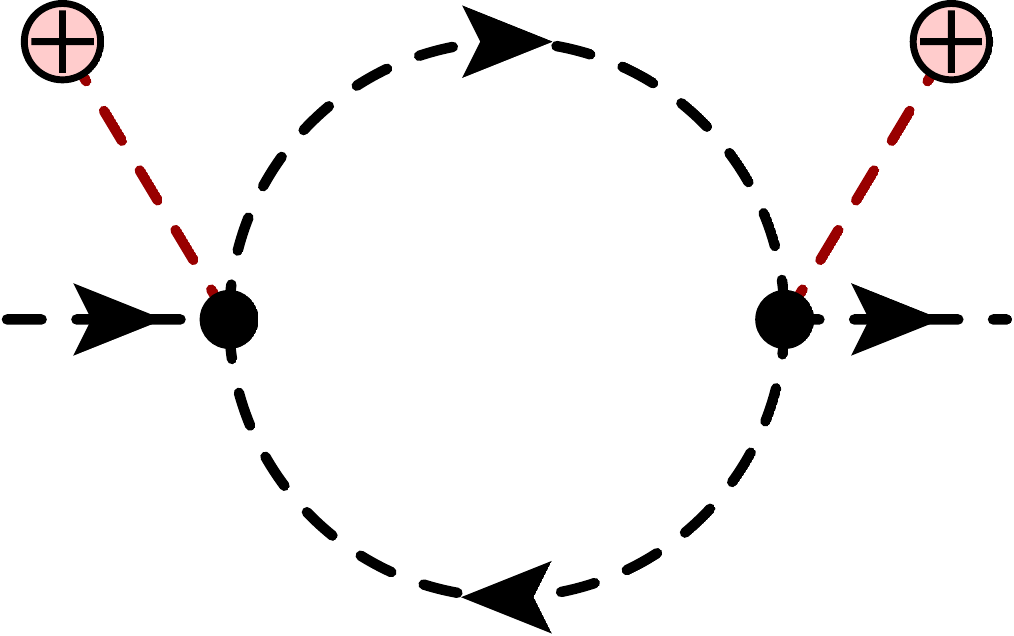}} 
      };
\node[anchor=east] at (-1.,0.25) {$p$};
\node[anchor=north] at (0.,1.4) {$k+p$};
\node[anchor=south] at (0.,-1.33) {$k$};
\node[anchor=east] at (-1.,1.2) {{\color{rossos}{$\phi$}}};
\node[anchor=east] at (1.55,1.2) {{\color{rossos}{$\phi$}}};
\node[anchor=east] at (-1.,-0.4) {$h$};
\node[anchor=west] at (.9,-0.4) {$h$};
\end{tikzpicture}
}}
=  -\frac{2M^4_{h}}{\phi^2}\int\frac{d^4k}{(2\pi)^4}P_{h}(p+k)P_{h}(k) = \frac{2iM^4_{h}}{\phi^2(4\pi)^2}
B_0(p^2,M^2_{h},M^2_{h})\,,\label{eq:ScalarHiggsLopop}
\end{align} 
with $P_{h}(k) = i/(k^2 - M_h^2)$. 
We also have the analogue diagram with the Goldstone $G$ running in the loop that gives a contribution with the same analytical form but $M_h^2 \to M_G^2 = (\lambda + \xi g^2/4)\phi^2$. 
Using eq.\,(\ref{eq:B0Der}), we find
\begin{align}
\left.Z_{\rm eff}^{(1)}\right|_{\rm Higgs} & =  \frac{\kappa M^2_{h}}{3\phi^2} = \kappa \lambda\,,\label{eq:ZHiggs}\\
\left.Z_{\rm eff}^{(1)}\right|_{\rm Goldstone} & =\frac{\kappa M^2_{G}}{3\phi^2} =\frac{\kappa}{3}\left(\lambda + \frac{\xi g^2}{4}\right)\,.\label{eq:ZGoldstone}
\end{align}
The ghost contribution is
  \begin{align}
 \left.i\Sigma_{hh}^{(1)}\right|^{\rm ghosts}
 &=
\resizebox{25mm}{!}{
\parbox{30mm}{
\begin{tikzpicture}[]
\node (label) at (0,0)[draw=white]{ 
       {\fd{2.75cm}{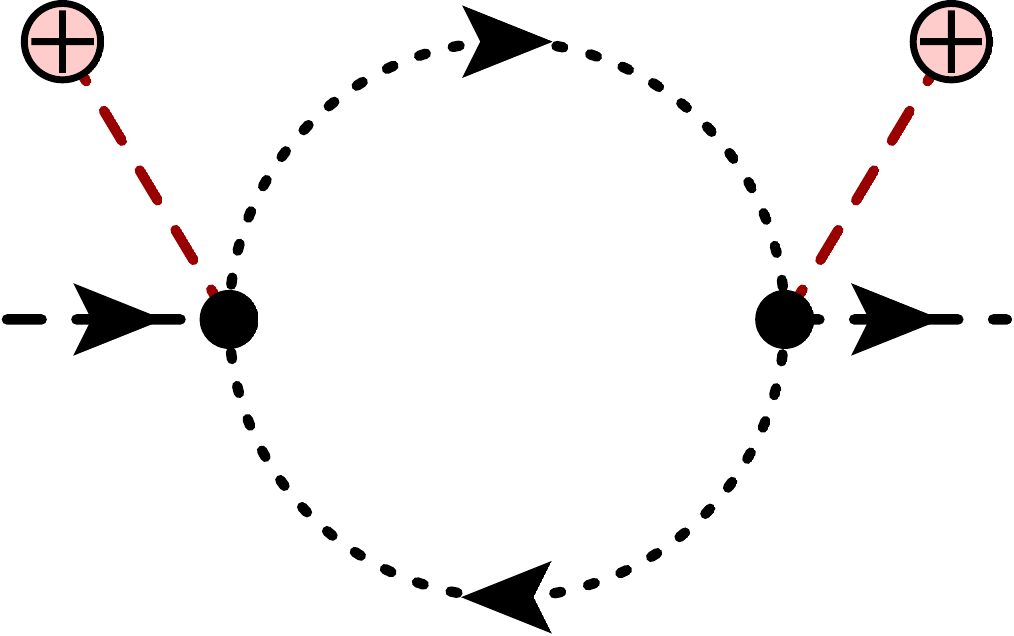}} 
      };
\node[anchor=east] at (-1.,0.25) {$p$};
\node[anchor=north] at (0.,1.4) {$k+p$};
\node[anchor=south] at (0.,-1.33) {$k$};
\node[anchor=east] at (-1.,1.2) {{\color{rossos}{$\phi$}}};
\node[anchor=east] at (1.55,1.2) {{\color{rossos}{$\phi$}}};
\node[anchor=east] at (-1.,-0.4) {$h$};
\node[anchor=west] at (.9,-0.4) {$h$};
\end{tikzpicture}
}} = +\frac{4M_c^4}{\phi^2}\int\frac{d^4k}{(2\pi)^4}P_{c}(p+k)P_{c}(k) =
 -\frac{i4M^4_{c}}{\phi^2(4\pi)^2}
B_0(p^2,M^2_{c},M^2_{c})\,,
\end{align} 
where the overall sign changes with respect to eq.\,(\ref{eq:ScalarHiggsLopop}) due to the Grassmannian nature of ghost fields (and there is no symmetry factor $1/2$).
We find
\begin{align}
\left.Z_{\rm eff}^{(1)}\right|_{\rm ghosts} & =  -\frac{2\kappa M^2_{c}}{3\phi^2} = - \frac{\kappa \xi g^2}{6} \,.\label{eq:ZGhost}
\end{align}
Finally, we have the mixing contribution 
  \begin{align}
 \left.i\Sigma_{hh}^{(1)}\right|^{\rm mix}
 &=
\resizebox{25mm}{!}{
\parbox{30mm}{
\begin{tikzpicture}[]
\node (label) at (0,0)[draw=white]{ 
       {\fd{2.75cm}{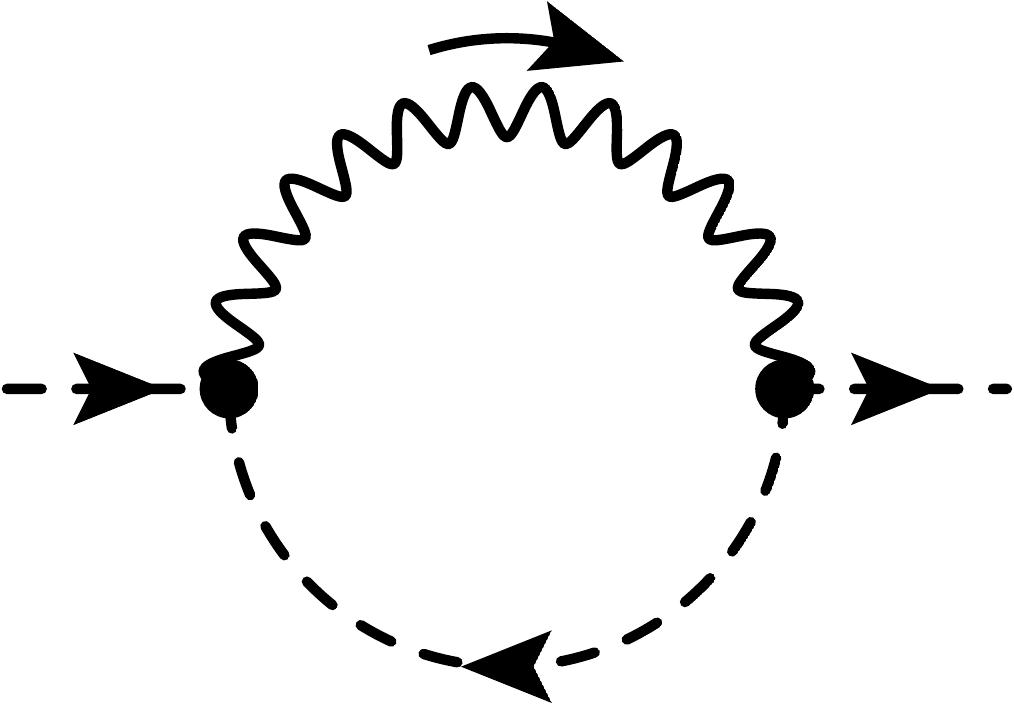}} 
      };
\node[anchor=east] at (-1.,0.25) {$p$};
\node[anchor=north] at (0.5,1.5) {$k+p$};
\node[anchor=south] at (-0.2,-1.33) {$k$};
\node[anchor=east] at (-1.,-0.4) {$h$};
\node[anchor=west] at (.9,-0.4) {$h$};
\end{tikzpicture}
}} =  -g^2 p^{\mu}p^{\nu}\int\frac{d^4 k}{(2\pi)^4}P_{\mu\nu}^{(\xi)}(p+k,M_A)P_{G}(k)\nn\\
&= \frac{ig^2}{4(4\pi)^2 M_A^2}\times \\
&\hspace{1cm}\left\{
(1-\xi)M_A^2A_0(M_{G}^2) + 
\left(M_{G}^2 - M_A^2 - p^2\right)A_0(M_A^2) + \left(p^2-M_{G}^2 - \xi M_A^2\right)A_0(\xi M_A^2)
\right.\nn\\
&\hspace{1.15cm}\left.
+\left[
M_{G}^4 - 2M_{G}^2M_A^2 + M_A^4 -2p^2\left(M_{G}^2 + M_A^2\right) + p^4
\right]B_0(p^2,M_A^2,M_{G}^2)
\right.\nn\\
&\hspace{1.15cm}\left.
-\left[
M_{G}^4 - 2\xi M_{G}^2M_A^2 + \xi^2M_A^4 -2p^2\left(M_{G}^2 -\xi M_A^2\right) + p^4
\right]B_0(p^2,M_{G}^2,\xi M_A^2)
\right\}\,,
\end{align} 
that follows from eq.\,(\ref{eq:MixingInteractions}) if the background field is non-constant.
By simple power counting, we see that this diagram is UV divergent. As in the rest of the paper, we implement the ${\overline{{\rm MS}}}$ renormalization scheme, and we find
 the renormalized quantity
\begin{align}\label{eq:Zmix}
\left. Z^{(1)}_{\rm eff}\right|_{\rm mix} = \frac{\kappa g^2}{8}\bigg[
-&5 - 3\xi - \frac{2M_G^2}{(M_A^2 -M_G^2)}\log\frac{M_A^2}{M_G^2} + \frac{2\xi M_G^2}{(M_G^2 - M_c^2)}\log\frac{M_G^2}{M_c^2}\nn\\
+& \frac{8 M_G^2}{(M_G^2 - M_A^2)}\log\frac{M_G^2}{\mu^2} + 2\xi\log\frac{M_c^2}{\mu^2} + \frac{2(3M_A^2 + M_G^2)}{(M_A^2 - M_G^2)}\log\frac{M_A^2}{\mu^2}\bigg]\,.
\end{align}

\subsection{Relevant integrals}
\label{app:B}

We use dimensional regularization in $d=4-2\epsilon$ dimensions.
The scalar one-point integral is
\begin{equation}
A_0(m^2) = \frac{(2\pi\mu)^{4-d}}{i\pi^2}\int d^dk \frac{1}{k^2 - m^2} =m^2\left(\Delta_{\rm UV} - \log\frac{m^2}{\mu^2} + 1 \right)\,,
\end{equation}
with $\Delta_{\rm UV} \equiv  1/\epsilon - \gamma_{\rm E} +\log(4\pi)$. The scalar two-point integral is
\begin{align}
B_0(p^2,m_1^2,m_2^2) &= \frac{(2\pi\mu)^{4-d}}{i\pi^2}\int d^dk\frac{1}{(k^2-m_1^2)[(k+p)^2 - m_2^2]}\nn\\
&=\Delta_{\rm UV} + \log\mu^2 -\int_0^1dx\,\log\left[
xm_1^2 + (1-x)m_2^2 -x(1-x)p^2
\right]\,.\label{eq:UsefulB}
\end{align}
We find the following limits used in appendix\,\ref{app:A}.
\begin{eqnarray}
B_0(0,m^2,m^2) &=&\Delta_{\rm UV} - \log\frac{m^2}{\mu^2}\,,\\
B_0(0,m_1^2,m_2^2) &=& \Delta_{\rm UV} + 1 - \frac{m_1^2}{m_1^2 - m_2^2}\log\frac{m_1^2}{\mu^2} + \frac{m_2^2}{m_1^2 - m_2^2}\log\frac{m_2^2}{\mu^2}\,,\\ 
B_0(0,0,m^2) &=& \Delta_{\rm UV} +1 - \log\frac{m^2}{\mu^2}\,,\\
\left.\frac{d}{dp^2}B_0(p^2,m_1^2,m_2^2)\right|_{p^2 = 0} &=& \frac{m_1^2 + m_2^2}{2(m_1^2 - m_2^2)^2} 
-\frac{m_1^2 m_2^2}{(m_1^2 - m_2^2)^3}\log\frac{m_1^2}{m_2^2}\,,\\
\left.\frac{d}{dp^2}B_0(p^2,m^2,m^2)\right|_{p^2 = 0} &=& \frac{1}{6m^2}\,,\label{eq:B0Der}\\
\left.\frac{d}{dp^2}B_0(p^2,0,m^2)\right|_{p^2 = 0} &=&\frac{1}{2m^2}\,.
\end{eqnarray}




\end{document}